\documentclass[twocolumn,superscriptaddress,aps,preprintnumbers,amsmath,amssymb,sort&compress,pra,nofootinbib]{revtex4-1}

\usepackage{amsmath}
\usepackage{amssymb}
\usepackage{mathtools}
\usepackage{slashed}
\usepackage{subfigure}
\usepackage{cancel}
\usepackage{textcomp}
\usepackage{calc}
\usepackage{graphicx}
\usepackage{dcolumn}
\usepackage{color}
\usepackage{colortbl}
\usepackage{bm}
\usepackage[utf8]{inputenc}
\usepackage{comment}
\usepackage{ulem}

\usepackage{color}
\definecolor{cred}{RGB}{180,50,40} 
\definecolor{purple}{RGB}{180,90,180} 
\definecolor{darkgreen}{RGB}{0, 100, 0}
\definecolor{dark_red}{rgb}{0.7, 0., 0.}
\definecolor{light_pink}{rgb}{1,0.4,0.4}
\definecolor{light_blue}{rgb}{0.284602,0.317763,0.963947}

\usepackage[colorlinks=true,urlcolor=dark_red,linkcolor=light_blue,citecolor=light_pink]{hyperref}

\usepackage{fourier}

\newcommand{\vev}[1]{ \left\langle {#1} \right\rangle }

\newcommand{\GeV}{\ \text{GeV} }

\newcommand{\dd}{\mathrm{d}}
\newcommand{\Mpl}{M_{\text{Pl}}}
\newcommand{\abs}[1]{\left\vert {#1} \right\vert}

\begin{document}

%%%%%%%%%%%%%%%%%%%%%%%%%%%%%%%%%%%%%%%%%%%%%%%%%%%%%%%%%%%%%%%%%%%%%%%%%%%%%%%%%%%%%%%%%%%%%%%%%%%%

\title{Baryogenesis from axion inflation}

\author{Valerie Domcke}
\affiliation{Deutsches Elektronen-Synchrotron (DESY), 22607 Hamburg, Germany}

\author{Benedict von Harling}
\affiliation{Deutsches Elektronen-Synchrotron (DESY), 22607 Hamburg, Germany}
\affiliation{IFAE, Universitat Aut\`onoma de Barcelona, 08193 Bellaterra (Barcelona), Spain}

\author{Enrico Morgante}
\affiliation{Deutsches Elektronen-Synchrotron (DESY), 22607 Hamburg, Germany}

\author{Kyohei Mukaida}
\affiliation{Deutsches Elektronen-Synchrotron (DESY), 22607 Hamburg, Germany}

\preprint{DESY 19-084}

%%%%%%%%%%%%%%%%%%%%%%%%%%%%%%%%%%%%%%%%%%%%%%%%%%%%%%%%%%%%%%%%%%%%%%%%%%%%%%%%%%%%%%%%%%%%%%%%%%%%

\begin{abstract}
\noindent
The coupling of an axion-like particle driving inflation to the Standard Model particle content through a Chern-Simons term generically sources a dual production of massless helical gauge fields and chiral fermions. We demonstrate that the interplay of these two components results in a highly predictive baryogenesis model, which requires no further ingredients beyond the Standard Model. If the helicity stored in the hyper magnetic field and the effective chemical potential induced by the chiral fermion production are large enough to avoid magnetic diffusion from the thermal plasma but small enough to sufficiently delay the chiral plasma instability, then the non-vanishing helicity survives until the electroweak phase transition and sources a net baryon asymmetry which is in excellent agreement with the observed value. If any of these two conditions is violated, the final baryon asymmetry vanishes. 
The observed baryon asymmetry can be reproduced if the energy scale of inflation is around $H_\text{inf} \sim 10^{10}$--$10^{12}$\, GeV with a moderate dependence on inflation model parameters.

\end{abstract}

%%%%%%%%%%%%%%%%%%%%%%%%%%%%%%%%%%%%%%%%%%%%%%%%%%%%%%%%%%%%%%%%%%%%%%%%%%%%%%%%%%%%%%%%%%%%%%%%%%%%

\date{\today}
\maketitle

%%%%%%%%%%%%%%%%%%%%%%%%%%%%%%%%%%%%%%%%%%%%%%%%%%%%%%%%%%%%%%%%%%%%%%%%%%%%%%%%%%%%%%%%%%%%%%%%%%%%

\section{Introduction}

Explaining today's evident asymmetry between matter and antimatter requires a tiny asymmetry between baryons and antibaryons in the primordial plasma, reflected by the present day ratio of the net number of baryons  and the entropy density, $\eta_B \simeq 9 \times 10^{-11}$~\cite{Tanabashi:2018oca}. In this paper we argue that this small number may be the generic outcome of any cosmic inflation model invoking large enough shift-symmetric couplings to the Standard Model (SM) particle content. The coupling of an inflaton $\phi$ to the Chern-Simons term of the SM hypercharge gauge group~$U(1)_Y$, $\phi Y_{\mu \nu} \tilde Y^{\mu \nu}$ (which up to total derivatives respects the shift symmetry $\phi \mapsto \phi + c$) leads to rapid non-perturbative production of helical gauge fields~\cite{Turner:1987bw, Garretson:1992vt, Anber:2006xt,Caprini:2014mja,Adshead:2016iae,Caprini:2017vnn}. Under certain conditions, these long-range helical hyper magnetic fields survive until the electroweak (EW) phase transition. The anomaly of baryon plus lepton number, $B+L$, links changes in these helical hyper magnetic fields to changes in the total $B+L$-charge. 
At the EW phase transition, the hyper magnetic field is converted into an electromagnetic field. Since the latter does not contribute to the anomaly, a compensating $B+L$-asymmetry is generated in this process. However, EW sphalerons stay in thermal equilibrium until the temperature reaches $T\simeq 130 \GeV$ and threaten to (almost) completely erase this asymmetry. By carefully studying the dynamics at the EW crossover, Ref.~\cite{Kamada:2016cnb} (see \cite{Joyce:1997uy,Bamba:2006km,Kamada:2016eeb} for important earlier work) has shown that this does not happen and that a sizable net baryon asymmetry remains.
This mechanism has recently been studied in Ref.~\cite{Jimenez:2017cdr} (see also \cite{Anber:2015yca} for earlier work), demonstrating that the correct baryon asymmetry can be reproduced for the appropriate parameter choices.

Here we extend Ref.~\cite{Jimenez:2017cdr} to include the effects of the SM fermions, which will have a significant impact on the final baryon asymmetry. The SM chiral anomaly ensures that during inflation, helical gauge fields \textit{and} chiral fermions are produced simultaneously~\cite{Domcke:2018eki,Domcke:2018gfr}, respecting in particular the chiral anomaly equation. 
Namely, starting from zero asymmetries for both charges, we expect that\footnote{
	Here we assume that the sphaleron and Yukawa interactions are not efficient during inflation.
}
\begin{align}
 n_\alpha - n_{\bar \alpha} = - \frac{\epsilon_\alpha}{2}  N_\alpha Q_{Y, \alpha}^2 q_{_\text{CS,Y}} \,,
 \label{eq:anomalyIntro}
\end{align}
where the index $\alpha$ runs over all SM fermion species with hypercharge $Q_{Y, \alpha}$ and multiplicity $N_\alpha$; $n_\alpha$ ($n_{\bar \alpha}$) denotes the number density of (anti-)particles and $\epsilon_\alpha = \pm$ accounts for right-/left-handed fermions. The Chern-Simons charge $q_{_\text{CS,Y}}$ is directly related to the helicity $h$ of the hyper gauge fields,
 $q_{_\text{CS,Y}} = (\alpha_Y/\pi) h$ with $h = (\text{vol}\, (\mathbb{R}^3))^{-1} \int \dd^3 x\, \bm{A}_Y \cdot \bm{B}_Y$ where $\bm{A}_Y$ ($\bm{B}_Y$) denote the hypercharge vector potential (hyper magnetic field) and $\alpha_Y$ is the corresponding fine-structure constant.
Once a thermal plasma forms after the end of inflation, the net number density can be expressed in terms of a chemical potential, $n_\alpha - n_{\bar \alpha} = N_\alpha \mu_\alpha T^2/6$.\footnote{We have expanded in $\mu_\alpha/T$ for $\mu_\alpha/T \ll 1$ which will always be fulfilled in the cases of interest.} 
We will be particularly interested in the right-handed electron which is the most weakly coupled SM particle to the sphaleron (through its Yukawa coupling to the left-handed electron), with
\begin{align}
  \mu_e T^2 = - 3 q_{_\text{CS,Y}} \,,
  \label{eq:mue}
\end{align}
because the asymmetry stored in the right-handed electron  survives the longest.
The task is now to track the evolution of the helical hyper gauge fields throughout the radiation dominated Universe until the EW phase transition in the presence of non-zero chemical potentials for all SM particles.
A direct consequence of these chemical potentials is the chiral plasma instability~\cite{Joyce:1997uy, Boyarsky:2011uy, Akamatsu:2013pjd,Hirono:2015rla, Yamamoto:2016xtu, Rogachevskii:2017uyc, Kamada:2018tcs} (see also~\cite{Brandenburg:2017rcb,Schober:2017cdw,Schober:2018ojn}) induced by the chiral magnetic effect~\cite{Vilenkin:1980fu, Alekseev:1998ds, Son:2004tq, Fukushima:2008xe}. In the absence of chemical potentials and for a sufficiently large electric conductivity of the SM plasma, the helicity stored in the hyper magnetic field is an approximately conserved quantity. Inverse cascade phenomena, induced by the velocity field of the SM plasma, ensure that the helicity is transferred from small scales to larger scales, thereby protecting it from competing diffusion effects~\cite{Pouquet:1976zz, Kahniashvili:2012uj,Banerjee:2004df}. However, the presence of non-vanishing chemical potentials with the opposite sign compared to the Chern-Simons charge [as induced \textit{e.g.}, by Eq.~\eqref{eq:mue}] causes the enhancement of modes with opposite helicity, thus threatening to erase the net helicity of the system. 

In order to allow for a non-vanishing net helicity at the EW phase transition, we need to 
ensure that this erasure due to the chiral plasma instability does not occur. To this end, it is enough to require that the chiral plasma instability does not happen before the weakest coupled particle, the right-handed electron, enters into thermal equilibrium at a temperature of about $10^5$~GeV. After this, the chemical potentials of all SM particles approximately vanish due to the sphaleron processes. This requirement imposes an upper bound on the chemical potential of the right-handed electron and, via its connection to the Chern-Simons charge, on the energy scale of inflation. Similarly, in order to ensure that diffusion effects do not wash out the net helicity of the hyper gauge field, we need to require that 
{the thermal plasma develops a significant/non-trivial velocity field}
 and prevents diffusion. 
This imposes a lower bound on the gauge field production and hence on the energy scale of inflation. 
Combining these two requirements allows us to predict the baryon asymmetry of the Universe. If one of the two requirements is not met, the total baryon asymmetry is driven to zero. If both requirements are met, the baryon asymmetry remarkably comes out to have the right order of magnitude, with an uncertainty based mainly on current uncertainties in SM computations (such as the dynamics of the EW phase transition and the evolution of a chiral SM plasma in a helical gauge field background), with only a 
mild dependence on the only remaining relevant model parameter, which is the effective coupling governing the strength of the $\phi Y_{\mu \nu} \tilde Y^{\mu \nu}$ term.

{Two crucial ingredients in our analysis are the estimate of the time-scale of the chiral plasma instability and the criterion {to avoid the magnetic diffusion} from the SM thermal plasma. For the former, }
we extend the discussion of \cite{Boyarsky:2011uy} (see also \cite{Hirono:2015rla}) to include a chemical potential whose sign is opposite to $q_{_\text{CS,Y}}$, as indicated by Eq.~\eqref{eq:mue}. Starting from a nearly maximally helical gauge field configuration, we observe (beginning at small wavelengths and proceeding to larger wavelengths) that the helicity of each Fourier mode is decreased (in absolute value), switches sign and grows again (in absolute value), thriving to drive the net helicity of the system to zero. We hence find an inverse cascade behavior, with a maximally helical gauge field configuration in the presence of an opposite sign chemical potential forming an unstable fix point of the theory.  Our numerical studies confirm our analytical estimate of the temperature $T_\text{CPI}$ below which the chiral plasma instability becomes effective.

For the criterion to avoid magnetic diffusion, we point out that for the length scales relevant for primordially generated (hyper) magnetic fields, the kinetic Reynolds number of the SM thermal plasma is typically small. This threatens to erase the velocity field of the plasma, thus preventing the onset of a turbulent or viscous regime which is required to protect the hyper magnetic fields from diffusion. We perform an analytical estimate finding a fix-point of the coupled magnetohydrodynamics (MHD) equations describing the velocity and hyper magnetic field. The hyper magnetic field acts as a source term for the velocity field, thus yielding a non-zero solution for the velocity field. Requiring the resulting magnetic Reynolds number to be sufficiently large, thus preventing the diffusion of the hyper magnetic field, imposes a lower bound on the Hubble parameter in our parameter space. Although this criterion should be taken as an order of magnitude estimate only, it demonstrates that (i) the requirement to avoid the magnetic diffusion can pose a stringent constraint on primordial magnetogenesis models and (ii) the regime in which {the hyper magnetic field is most likely to survive from the diffusion} coincides with a predicted baryon asymmetry in agreement with observations.

The remainder of this paper is organized as follows. We begin by summarizing gauge field and fermion production during inflation in Sec.~\ref{sec:axioninflation}. This sets the initial conditions for the evolution of the chemical potentials and helical magnetic fields until the EW phase transition, described in Sec.~\ref{sec:evolution}. We compute the resulting baryon asymmetry in Sec.~\ref{sec:baryogenesis} while implications for present-day intergalactic magnetic fields and anisotropies in the cosmic microwave background (CMB) are discussed in Sec.~\ref{sec:furtherobs}. We conclude in Sec.~\ref{sec:conclusion}. 
{We relegate some more technical but important results to three appendices. In App.~\ref{sec:chemical} we summarize chemical equations associated with the SM interactions. In App.~\ref{sec:reynolds} we derive our criterion for the onset of magnetic diffusion in the SM plasma. Finally, App.~\ref{sec:cpi} is dedicated to a study of the chiral plasma instability, including a derivation of the corresponding equations.}

%%%%%%%%%%%%%%%%%%%%%%%%%%%%%%%%%%%%%%%%%%%%%%%%%%%%%%%%%%%%%%%%%%%%%%%%%%%%%%%%%%%%%%%%%%%%%%%%%%%%%%%%%%%%%%%%%%%%%%%%%%%%%%%%
\section{Axion inflation \label{sec:axioninflation}}
\paragraph*{Setup.} We consider cosmic inflation to be driven by an axion-like pseudoscalar particle $\phi$, characterized by shift-symmetric couplings to the SM degrees of freedom.
The action relevant to our following discussion is\footnote{
Note that the sign of the fermion coupling to the gauge field differs from the convention often used in quantum field theory. It agrees, on the other hand, with the usual convention in electrodynamics, where, \textit{e.g.}, the magnetic field is calculated as ${\bm B} = \nabla \times {\bm A}$. Both are related via $A_\mu \mapsto -A_\mu$ or $Q \mapsto -Q$.
} 
\begin{align}
	S = &  \int \dd^4 x \, \Bigg\{ \sqrt{-g}
	\left[
		\frac{g^{\mu\nu}}{2} \partial_\mu \phi \partial_\nu \phi - V (\phi)
	\right]
		- \frac{1}{4} Y_{\mu \nu} Y^{\mu\nu}  \nonumber  \\
		&  + \sum_\alpha \overline \psi_\alpha \left( i \partial \cdot \gamma - g_Y Q_{Y, \alpha} A_Y \cdot \gamma \right) \psi_\alpha
	+ \frac{\alpha_Y \phi}{4 \pi f_a} Y_{\mu \nu} \tilde Y^{\mu \nu}
    	\Bigg\}. \label{eq:setup_conf1}
\end{align}
Here $g_Y$, $A_{Y,\mu}$ and $Y_{\mu \nu}$ denote the gauge coupling, vector potential and field strength tensor of the SM gauge group U$(1)_Y$. The dual field strength tensor is obtained as $\tilde Y^{\mu \nu} = \epsilon^{\mu \nu \rho \sigma} Y_{\rho \sigma}/2$ with $\epsilon^{0123} = + 1$.\footnote{
Here we have omitted the non-abelian SM gauge groups for several reasons. Firstly, due to their non-linear interactions, their production during inflation is suppressed compared to the abelian gauge fields for similar effective couplings $\alpha/f_a$~\cite{Adshead:2012kp,Dimastrogiovanni:2012st,Dimastrogiovanni:2012ew,Adshead:2013qp,Adshead:2013nka,Domcke:2018rvv}. Note, however, that in the presence of massless fermions, the backreaction through the induced fermion current (see below) has a significant impact on the abelian gauge field production~\cite{Domcke:2018eki} whereas it is negligible for the non-abelian case~\cite{Domcke:2018gfr}.
Secondly, after the end of inflation, we expect the non-abelian gauge fields to thermalize quickly through the non-abelian gauge interactions, washing out any helicity stored in this sector. 
Thirdly, as we will demonstrate, in the conversion of the helical gauge fields into a baryon asymmetry, a crucial role is played by the right-handed electron (being the most weakly coupled SM particle), which is not charged under the non-abelian SM gauge groups.
Consequently the non-abelian SM gauge groups do not contribute significantly to the generation of the baryon asymmetry of the Universe, and we will ignore them for simplicity in the following.
}
The massless SM chiral fermions are denoted by $\psi_\alpha$ where $\alpha$ runs over $e_i, L_i, u_i, d_i, Q_i$ with the generation index being $i=1,2,3$. Note that an appropriate projection operator, $\mathcal P_\text{R/L} = (1 \pm \gamma_5)/2$, is implicit in the definition of $\psi_\alpha$.
We have dropped the SM Yukawa interactions since they play no role during inflation.
The coupling of the inflaton $\phi$ to the gauge fields and fermions is determined through the hypercharge fine-structure constant $\alpha_Y = g_Y^2/(4 \pi)$ and the axion decay constant $f_a$. The shift symmetry of the inflaton is (mildly) broken through the scalar potential $V(\phi)$, in accordance with slow-roll inflation. 

We will work mainly in the conformal frame, in which indices are raised and lowered by the Minkowski metric $\eta^{\mu \nu} = \text{diag}(1, -1, -1, -1)$ and $\{ \gamma^\mu , \gamma^\nu \} = 2 \eta^{\mu \nu}$. In particular, this is implicit in the dot-product notation of Eq.~\eqref{eq:setup_conf1}, \textit{e.g.}, $\partial \cdot \gamma = \eta^{\mu \nu} \partial_\mu \gamma_\nu$.
Similarly, the vector potential $A_{Y,\mu}$ and the fermions $\psi_\alpha$ in Eq.~\eqref{eq:setup_conf1} are comoving fields, from which the corresponding physical fields are obtained as 
$\hat A_Y^\mu = A_Y^\mu/a^2$, $\hat A_{Y,\mu} = A_{Y,\mu}$ and $\hat \psi_\alpha = a^{-3/2} \psi_\alpha$ with $a$ being the scale factor of the FLRW metric. The three-dimensional vector potential is defined as $A_Y^\mu = (A_{Y,0},{\bm A}_Y)$, $A_{Y,\mu} = (A_{Y,0},-{\bm A}_Y)$. Throughout this paper, we will denote physical quantities (as opposed to comoving ones) by a hat.
The expansion of the Universe, encoded in the corresponding FLRW metric $g_{\mu \nu} = a^2(t) \, \eta_{\mu \nu}$, hence appears in Eq.~\eqref{eq:setup_conf1} only in the kinetic term of the scalar and in the overall factor $\sqrt{-g} = [- \text{det}(g_{\mu \nu})]^{1/2}= a^4(t)$.

At first glance, the Chern-Simons coupling between the inflaton and  the hyper gauge field, given in Eq.~\eqref{eq:setup_conf1}, has nothing to do with the SM fermions.
However, the chiral anomaly relates each fermion current with the Chern-Simons density as follows:\footnote{Throughout this paper, we will employ Einstein's sum convention for Lorentz indices $\{\mu, \nu,..\}$ but not for the index $\alpha$ running over the SM particles.}
\begin{align}
	\partial_\mu J^\mu_{\alpha} = - \epsilon_\alpha N_\alpha \frac{g_Y^2 Q_{Y, \alpha}^2}{16 \pi^2}  Y_{\mu \nu} \tilde Y^{\mu \nu} + \cdots 
 \label{eq:anomaly}
\end{align}
Here the SM fermions $\alpha = e_i,L_i,u_i,d_i,Q_i$ appear with multiplicity $N_\alpha = 1,2,3,3,6$ and have hypercharge $Q_{Y, \alpha} = -1,-1/2,2/3,-1/3,1/6$, respectively.\footnote{
	$Q_i$ denotes the left-handed quark doublet for the $i$-th generation, which should not be confused with the hypercharge, $Q_{Y, \alpha}$.
}  We have suppressed the SU$(2)_W$, SU$(3)_C$, and Yukawa interactions as ellipsis.
After the appropriate summation over each species, this yields the well-known non-conservation of $B+L$ through non-perturbative processes. Note that $B+L$ is conserved in Yukawa interactions, implying an exact relation between the $B+L$ current and the Chern-Simons density.
In the context of axion inflation, Eq.~\eqref{eq:anomaly} implies a simultaneous production of gauge fields and massless fermions, which, as we briefly review in the following, results in the production of helical hyper gauge fields as well as SM chiral fermions~\cite{Domcke:2018eki}. We emphasize that this mechanism is complementary to the production of massive fermions during axion inflation discussed in \cite{Adshead:2015kza,Adshead:2018oaa}.

\vspace{1em}\paragraph*{Helical gauge field and chiral fermion production.} During inflation, the energy budget of the Universe is dominated by vacuum energy, $V(\phi) \gg V'(\phi) \Mpl, \dot \phi^2$. Neglecting for a moment any effects associated with the fermions, Eq.~\eqref{eq:setup_conf1} leads to the equation of motion for the gauge field in Fourier space,
\begin{align}
	0 = \left[ \partial_\eta^2 + k \left( k \pm 2 \lambda \xi a H \right) \right] A_{Y,\pm} (\eta, \bm{k}) \,,
	\label{eq:wave_eq}
\end{align}
where $\lambda = \pm$ for $\dot \phi \gtrless 0$ encodes the sign of $\dot \phi$,
\begin{align}
	\xi \equiv \frac{\alpha_Y { \lambda \dot \phi}}{2 \pi f_a H} > 0 \,,
	\label{eq:def_xi}
\end{align}
and we have expressed the gauge field in terms of the two helicity eigenstates with respect to the wave vector $\bm k$ of the Fourier mode in question. $H = \dot a/a$ is the Hubble parameter. $A_{Y,\pm}$ are the transverse mode functions of the field $A_Y^\mu$ and have the property $A_{Y,\pm} (\eta,{\bm k})^* = A_{Y,\pm} (\eta,-{\bm k})$, while the polarization vectors satisfy $\epsilon_{\bm k}^{\pm\mu} = (0,{\bm\epsilon}_{\bm k}^\pm)$, ${\bm k}\cdot{\bm\epsilon}_{\bm k}^\pm = 0$, $({\bm\epsilon}_{\bm k}^\sigma)^*\cdot{\bm\epsilon}_{\bm k}^{\sigma'} = \delta_{\sigma\sigma'}$, ${\bm k}\times{\bm\epsilon}_{\bm k}^\pm = \mp i k{\bm\epsilon}_{\bm k}^\pm$ and $({\bm\epsilon}_{\bm k}^\pm)^* = {\bm\epsilon}_{-{\bm k}}^\pm$.
Here and in the following a dot denotes the derivative with respect to cosmic time $t$, related to the conformal time coordinate $\eta$ by $\dd t = a \, \dd\eta$.
In this system, parity is spontaneously violated by the choice of sign of the inflaton velocity $\dot \phi$, leading to a tachyonic enhancement of one of the two helicity modes of the gauge field.%
\footnote{Obtaining a positive baryon asymmetry will in the end require $\dot \phi < 0$. The opposite case, $\dot \phi > 0$, corresponds to a $CP$ transformation, i.e.\ ${\bm B}_Y \mapsto - {\bm B}_Y$, ${\bm E}_Y \mapsto {\bm E}_Y$, which results in the opposite sign for the total helicity and the final baryon asymmetry. In the equations below we will introduce the absolute values of ${\bm B}_Y$ and ${\bm E}_Y$, which are invariant under $CP$. On the contrary, $CP$-odd quantities will depend on the sign of ${\bm E}_Y \cdot {\bm B}_Y$.}

In the limit of constant $\xi$ (consistent with the slow-roll approximation), the enhanced mode of Eq.~\eqref{eq:wave_eq} is given exactly by
\begin{align}
	A_{Y,- \lambda} (\eta , \bm{k}) = \frac{e^{\pi \xi / 2 }}{\sqrt{2 k}}
	W_{- \lambda i \xi,1/2} (2 i k \eta),
	\label{eq:gauge_desitter}
\end{align}
with $W_{k,m}(z)$ denoting the Whittaker function. For $\lambda = +1$ $(\lambda = -1)$ this leads to (anti-)parallel hyper electric and magnetic fields with a strongly enhanced amplitude on superhorizon scales. 

Due to the expansion of the Universe, the dominant contribution to the energy density in hyper gauge fields arises during the final e-folds of inflation. In the following, the parameter $\xi$ will therefore always be understood as being evaluated at this time. Note also that $\xi$ is not exactly constant during inflation and instead typically increases slowly. We will therefore not impose an upper bound on $\xi$ arising from the production of primordial density perturbations at CMB scales as this bound is only relevant for $\xi$ evaluated at earlier times during inflation. For phenomenological purposes, we will be interested in values $\xi \gtrsim 1$. Using that the kinetic energy of the inflation is bounded by its potential energy, we find $\xi \lesssim \sqrt{3} \alpha_Y M_{\rm Pl} / (\sqrt{2} \pi f_a)$. We then need $f_a \lesssim \alpha_Y M_{\rm Pl}$ for $\xi \gtrsim 1$. In the context of natural inflation, it is well known that the axion field excursion needs to be super-Planckian for successful inflation, which hence requires an axion with two hierarchically different decay constants. This can, for example, be achieved using the clockwork mechanism \cite{Kaplan:2015fuy,Choi:2015fiu}, axion monodromy~\cite{McAllister:2008hb,Silverstein:2008sg} and by aligning multiple axions~\cite{Liddle:1998jc,Dimopoulos:2005ac,Kim:2004rp} or can also arise from nearly conformal sectors \cite{Fonseca:2019aux}.

The background of strong quasi homogeneous gauge fields (on sub-horizon scales) leads to non-perturbative fermion production. This process was studied in detail in~\cite{Domcke:2018eki}, finding two distinct production channels: (i) pair production, similar to Schwinger production, symmetrically produces fermions and anti-fermions and (ii) asymmetric fermion production, associated with the chiral anomaly~\eqref{eq:anomaly}. 

To study the backreaction of the fermion production on the helical gauge field in a rigorous way, we have to solve the coupled equations of motion for the U$(1)_Y$ gauge field and the SM fermions simultaneously in the closed-time-path formalism, which is however beyond the scope of this paper. Instead, in the following, we sketch two ways to estimate its effect by balancing the production of U$(1)_Y$ gauge fields and their annihilation into SM fermions: the first argument provides an upper bound on the gauge field on superhorizon scales, and the second one gives a self-consistent solution for the gauge field equation of motion in dynamical equilibrium.

The generation of these fermions leads to an induced current for the hyper gauge field,
\begin{align}
 J^\mu_Y & = \sum_\alpha Q_{Y, \alpha} \overline \psi_\alpha \gamma^\mu \psi_\alpha \,, 
\end{align}
which in turn inhibits the gauge field production. This current is orientated in the direction of the hyper electric field $\bm{E}_Y$ and, neglecting the scatterings among fermions and assuming that the production of fermions is much faster than the cosmic expansion, its amplitude  can be computed to be~\cite{Domcke:2018eki}:\footnote{
We assume that the Higgs field is stabilized at the origin during inflation by a positive Hubble-induced mass term {(arising, \textit{e.g.}, via $\zeta R |H|^2$ or $c |H|^2 V(\phi) / \Mpl^2$), which is heavy enough to suppress its Schwinger effect.
If (depending on the sign of $\zeta$ and $c$), these couplings instead provide a tachyonic Hubble-induced mass, the EW symmetry is broken during inflation, which will be investigated elsewhere.}
Furthermore, we also assume that the fermions are dominantly generated by the hyper gauge field during inflation as opposed to, \textit{e.g.}, by the non-abelian SM gauge fields~\cite{Domcke:2018gfr}.
}
\begin{align}
\label{eq:induced_noscat}
	\frac{1}{a^3}
	g_Y  \vev{|J_Y|}
	&\simeq \sum_\alpha N_\alpha
	\frac{\left( g_Y \abs{Q_{Y, \alpha}} \right)^3}{12 \pi^2}
		\coth \left( \frac{\pi \hat B_Y}{\hat E_Y} \right) 
	\hat{E}_Y  \hat{B}_Y \frac{1}{H} \,,
\end{align}
where $\langle \bullet \rangle$ denotes the expectation value.
For later convenience, we have introduced the physical hyper gauge fields $\hat{\bm{E}}_Y = {\bm E}_Y/a^2$, $\hat{\bm{B}}_Y = {\bm B}_Y/a^2$ with ${\bm E}_Y = - \partial_\eta {\bm A}_Y$ and ${\bm B}_Y = \nabla \times {\bm A}_Y$ in temporal gauge $A_{0,Y}=0$.
Note that here and below, the hyper gauge fields not written in boldface, $\hat E_Y$ and $\hat B_Y$, are averaged quantities, $\hat E_Y = \langle \hat{\bm{E}}_Y^2 \rangle^{1/2}$ and $\hat B_Y = \langle \hat{\bm{B}}_Y^2 \rangle^{1/2}$,
which represent their typical amplitudes in any given Hubble patch.
The evolution of the gauge field energy density $\rho_{A_Y}$ is governed by the equation \cite{Domcke:2018eki}
\begin{align}
	\dot {\hat \rho}_{A_Y} = - 4 H \hat\rho_{A_Y} + 2 \xi H \hat{E}_Y \hat{B}_Y
	- \hat{E}_Y 
	g_Y \vev{\abs{\hat{J}_Y}} 
	\,,
\end{align}
where $\hat{\bm{J}}_Y = \bm{J}_Y / a^3$ is the physical current.
The amplitudes of the hyper gauge fields in equilibrium ($\dot \rho_{A_Y} = 0$) must hence obey the algebraic equation
\begin{align}
\label{eq:curve}
	0 = - 2 H \left( \hat E_Y^2 + \hat B_Y^2 \right) + 2 \xi_\text{eff} H \hat{E}_Y \hat{B}_Y \,,
\end{align}
where
\begin{align}
\label{eq:xi_eff}
	\xi_\text{eff} = \xi - \sum_\alpha N_\alpha \frac{\left( g_Y \abs{Q_{Y, \alpha}} \right)^3}{24 \pi^2} 
	\coth \left( \frac{\pi \hat B_Y}{\hat E_Y} \right) \frac{\hat E_Y}{H^2}\,.
\end{align}

\begin{figure}
\centering
 \includegraphics[width = 0.9 \columnwidth]{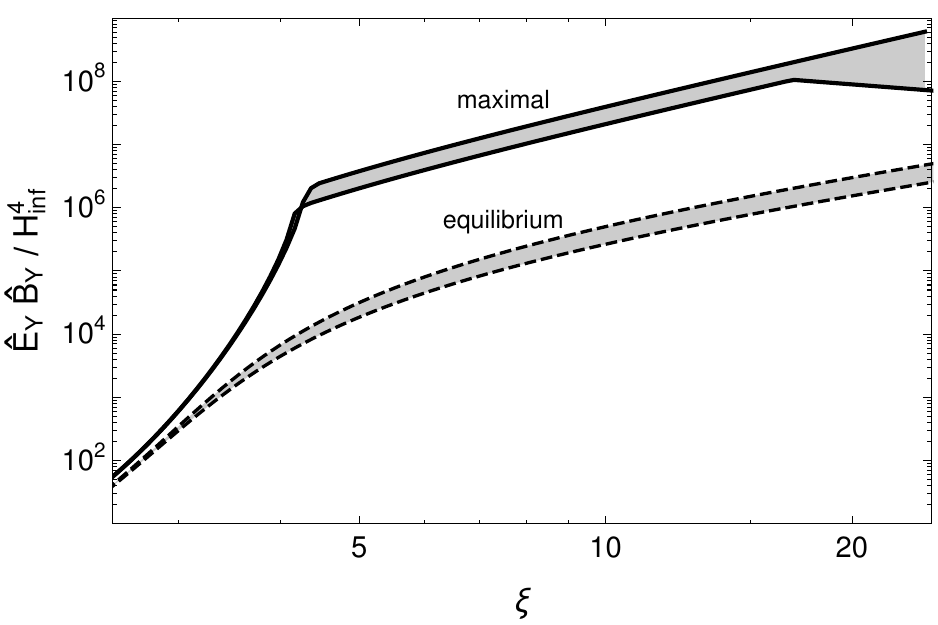}
 \caption{Helical gauge fields generated during inflation taking into account the backreaction from the induced fermion current. The solid lines show the upper bound and the dashed lines correspond to the equilibrium solution. In both cases, the gray band indicates the variation of the Hubble parameter in the range  $10^8~\text{GeV} \leq H_\text{inf} \leq 10^{14}~\text{GeV}$.}
 \label{fig:EB}
\end{figure}

Eq.~\eqref{eq:curve} gives an upper limit for the generated gauge fields, depicted by the solid lines in Fig.~\ref{fig:EB}. For $\xi \lesssim 4$ this coincides with the gauge field production described by Eq.~\eqref{eq:gauge_desitter}, for larger values of $\xi$ the backreaction of the induced fermion current becomes important, limiting the gauge field production. For $\xi \gtrsim 15$, the energy converted from the inflation sector to the hyper gauge and SM fermion sector per Hubble time becomes comparable to the vacuum energy driving inflation for large values of $H_\text{inf}$. In this regime, the gauge fields cannot consistently saturate the upper bound given by Eq.~\eqref{eq:curve} and their upper bound is instead determined by demanding that the gauge field and fermion production does not consume all the energy from the inflation sector within one Hubble time.
Therefore, although this bound is robust, we do not expect that it is saturated in reality because a significant backreaction on the inflaton dynamics is expected.

In addition to the upper bound, we can obtain a rough estimate of the magnitude of the produced gauge fields as follows:
If we assume that $\hat E_Y$ and $\hat B_Y$ are rapidly driven to their equilibrium solution, we may take $\xi_\text{eff}$ to be approximately constant and require that Eq.~\eqref{eq:gauge_desitter} [or Eq.~\eqref{eq:wave_eq}] with $\xi \mapsto \xi_\text{eff}$ describes the gauge field production in the presence of the fermion backreaction. The self-consistent choice of $\hat E^\text{eq}_Y$ and $\hat B^\text{eq}_Y$ meeting this requirement is shown as dashed lines in Fig.~\ref{fig:EB}. The gray bands in Fig.~\ref{fig:EB} correspond to the variation of the Hubble parameter in the range  $10^8~\text{GeV} \leq H_\text{inf} \leq 10^{14}~\text{GeV}$, entering through the renormalization of the gauge coupling constant,\footnote{
	Throughout this paper, we assume no new particles beyond the SM below the inflation scale. Note that a possible origin of the Chern-Simons coupling may involve new hypercharged particles, which however should be heavier than the inflation scale for a consistent usage of the effective field theory.
}
\begin{align}
 g_Y^{-2}(\mu) = g_Y^{-2}(m_Z) + \frac{41}{48 \pi^2} \ln \left( \frac{m_Z}{\mu} \right) \,,
\end{align}
evaluated here at the energy scale $\mu = (\tfrac{1}{2}(\hat E_Y^2 + \hat B_Y^2))^{1/4}$ with $g_Y(m_Z) \simeq 0.35$ at the $Z$-boson mass.
A larger Hubble parameter during inflation leads to a (slightly) larger gauge coupling and hence a stronger backreaction, reducing the amount of $\hat E_Y$ and $\hat B_Y$ fields produced.

\vspace{1em}\paragraph*{Approximately conserved charges.} 
The anomaly equation~\eqref{eq:anomaly} can be recast as a conservation equation linking the chiral fermion current $J_\alpha^\mu$ with the helical gauge field configuration sourced by the Chern-Simons term,
\begin{align}
	0 &= \partial_\mu \left( J_\alpha^\mu + \epsilon_\alpha \frac{N_\alpha Q_{Y, \alpha}^2}{2} J_{\text{CS},Y}^\mu \right) + \cdots\,,
	\label{eq:anomalous}
\end{align}
with $J_{\text{CS},Y}^\mu  \equiv (\alpha_Y/\pi) \epsilon^{\mu\nu\rho\sigma} A_{Y,\nu} \partial_\rho A_{Y,\sigma}$. 
This equation implies an approximate conserved quantity as long as the EW and strong sphalerons and the Yukawa interactions are not efficient, as is the case during inflation.
Let us assume this for a while. See the next section for the impacts of these interactions.
Starting from initial conditions with vanishing Chern-Simons/fermion charges, one finds that each individual fermion charge $q_\alpha$ and the Chern-Simons charge $q_{\text{CS},Y}$ are always balanced during axion inflation,
\begin{align}
\label{eq:charges anomaly}
 q_\alpha = - \epsilon_\alpha  \frac{N_\alpha Q^2_\alpha}{2} q_{\text{CS},Y} \,,
\end{align}
where the charge densities are obtained from the corresponding currents $J_\alpha^\mu$ and $J_{\text{CS},Y}^\mu$ as 
\begin{align}
	q_\bullet \equiv \frac{1}{\text{vol}\, (\mathbb{R}^3)} \int \dd^3 x\, \vev{ J_\bullet^0 }\,.
	\label{eq:charge}
\end{align}
We will sometimes use the helical charge $h$ instead of $q_{\text{CS},Y}$, related by $q_{\text{CS},Y} = (\alpha_Y / \pi) h$.
The charges $q_\alpha = n_\alpha - n_{\bar \alpha}$ account for the asymmetry between the number densities of particles and antiparticles for each fermion species.
In particular, the total $B+L$ asymmetry is obtained as
\begin{align}
	q_{B+L} = - \frac{3}{2} q_{\text{CS},Y}.
\end{align}
Here the factor $3$ comes from the three generations.
Thus, the value of all the charges $q_\alpha$ at the end of inflation can be expressed in terms of a single parameter, the Chern-Simons charge $q_{\text{CS},Y}$.

To estimate $q_{\text{CS},Y}$, we assume that the amplitudes of the physical hyper gauge fields become approximately constant on the time-scale of a Hubble time $H^{-1}$,
namely that the production of helical gauge fields and their decay via the cosmic expansion and the fermion production are balanced, as it happens without fermion backreaction. The evolution of the Chern-Simons charge during inflation is governed by the equation
\begin{align}
 	\partial_t q_{\text{CS},Y} = - a^3 \frac{2 \alpha_Y}{\pi} \langle \hat{\bm E}_Y \cdot \hat{\bm B}_Y \rangle\,,
	\label{eq:cs-evolution}
\end{align}
which we integrate under the assumption that $\langle \hat{\bm E}_Y \cdot \hat{\bm B}_Y \rangle \simeq \text{const.}$ and $\xi \simeq \text{const.}$ in a quasi-de-Sitter background, $a(t) \propto \exp(H t)$. Due to the exponential growth of the scale factor, the integral is dominated by the upper integration bound, so that the total Chern-Simons charge is determined by the value  $\langle \hat{\bm E}_Y \cdot  \hat{\bm B}_Y \rangle$ at the end of inflation. Assuming instantaneous reheating,\footnote{{As long as $\dot \phi \neq 0$, the non-perturbative gauge field production induced by the $\phi Y_{\mu \nu} \tilde Y^{\mu \nu}$ term efficiently converts the potential and kinetic energy of $\phi$ into radiation, both during inflation and in the subsequent (p)reheating period.  
In a lattice simulation (including axion and gauge fields), Ref.~\cite{Cuissa:2018oiw} (see also Ref.~\cite{Adshead:2015pva}) demonstrated preheating to be efficient, \textit{i.e.}, $\gtrsim 80 \%$ of the total energy is converted to radiation within about one e-fold after the end of inflation, for $\alpha_Y/f_a \gtrsim 10/\Mpl$. 
Neglecting the backreaction from the fermion current, this corresponds to $\xi(N = 1) \gtrsim 4$, evaluated one e-fold before the end of inflation, and hence (see Fig.~\ref{fig:EB}) $\hat E_Y \hat B_Y/H_\text{inf}^4 \gtrsim 2 \times 10^5$.
From this we conclude that the approximation of instant reheating is justified for $\xi \gtrsim 4$ if the gauge fields saturate their upper bound and for $\xi \gtrsim 9$ if the gauge fields are given by the equilibrium solution. For smaller values of $\xi$, the assumption of instantaneous reheating may induce significant systematic uncertainties related to the evolution of the Chern-Simons charge $q_{_\text{CS,Y}}$ over the course of reheating.
\label{ft:reheating}}
} this point in time coincides with the starting point of the subsequent radiation dominated phase which we denote with the subscript `rh'. The total Chern-Simons charge at the end of inflation then reads
\begin{align}
\frac{q^\text{rh}_{\text{CS},Y}}{a^3_\text{rh}} \simeq -\frac{2}{3} \frac{\alpha_Y}{\pi  H_\text{rh}} \langle \hat{\bm E}_Y \cdot \hat{\bm B}_Y \rangle_\text{rh} \,,
\label{eq:qcs0}
\end{align}
where, under the instantaneous reheating approximation, $H_\text{rh}$ coincides with $H_\text{inf}$ which denotes the (approximately constant) Hubble parameter during inflation.

%%%%%%%%%%%%%%%%%%%%%%%%%%%%%%%%%%%%%%%%%%%%%%%%%%%%%%%%%%%%%%%%%%%%%%%5

\section{Evolution of charges after inflation \label{sec:evolution}} 

In this section we discuss the evolution of the helical hyper gauge fields and the SM fermions in the radiation dominated regime between the end of inflation and the EW phase transition. 
Contrary to during cosmic inflation, Yukawa interactions and EW and strong sphaleron processes become relevant,  re-shuffling the charges $q_\alpha$ among the SM fermions.
These fermion asymmetries determine the decay of the helicity stored in the hyper gauge fields through the so-called chiral plasma instability (CPI)~\cite{Joyce:1997uy, Boyarsky:2011uy, Akamatsu:2013pjd,Hirono:2015rla, Yamamoto:2016xtu, Rogachevskii:2017uyc, Kamada:2018tcs} induced by the chiral magnetic effect~\cite{Vilenkin:1980fu, Alekseev:1998ds, Son:2004tq, Fukushima:2008xe}. 
{If the non-linear terms in the equations governing the evolution of the velocity field of the plasma and the hyper magnetic field dominate,}
then the plasma of SM particles and gauge fields can enter a viscous or turbulent regime in which the helicity is conserved against the diffusion~\cite{Pouquet:1976zz, Kahniashvili:2012uj,Banerjee:2004df}, and eventually, it is converted into a baryon asymmetry in the EW phase transition~\cite{Fujita:2016igl, Kamada:2016eeb, Kamada:2016cnb}. The purpose of this section is to identify the conditions under which the generation of this baryon asymmetry is not spoiled by the chiral plasma instability or the diffusion.

\vspace{1em}
\paragraph*{Sphalerons and Yukawa interactions.}
Let us assume kinetic equilibrium due to the frequent interactions, so that the distribution functions can be well approximated by the Fermi-Dirac or Bose-Einstein one with chemical potentials.
Under this assumption, we can express each charge $q_\alpha$ in terms of a chemical potential $\mu_\alpha$,
\begin{align}
	q_\alpha = \frac{1}{6} N_\alpha \mu_\alpha T^2\,, \quad
	q_H = \frac{2}{3}  \mu_H T^2\,,
	\label{eq:rel_chempot_charge}
\end{align}
where besides the fermions $\psi_\alpha$, we have also included the SM Higgs $H$. 
Note that $\mu_\bullet$ and $T$ shown here are comoving quantities whose relations to the physical quantities are given by $\hat \mu_\bullet = \mu_\bullet / a$ and $\hat T = T / a$. In particular, in an adiabatically expanding Universe and for a constant number of degrees of freedom in the thermal bath, the conformal temperature $T$ is simply a constant reference temperature.

Once a given Yukawa-coupling-mediated or sphaleron process becomes efficient, \textit{i.e.}, once it becomes fast compared to the cosmic expansion, the chemical potentials of the particles involved are reshuffled and driven to their chemical equilibrium value.
The EW and strong sphalerons lead to
\begin{align}
	0 = \sum_i \left( 3 \mu_{Q_i} + \mu_{L_i} \right)\,,
	\quad
	0 = \sum_i \left( 2 \mu_{Q_i} - \mu_{u_i} - \mu_{d_i} \right)\,,
	\label{eq:sph}
\end{align}
respectively.
The Yukawa interactions for up-type and down-type quarks give
\begin{align}
	0 = \mu_{Q_i} + \mu_H - \mu_{u_j}\,, \quad
	0 = \mu_{Q_i} - \mu_H - \mu_{d_j}
	\label{eq:y_q}
\end{align}
for all pairs $i,j$. As follows from these relations, the chemical potentials of the quarks become generation independent, \textit{i.e.,} $\mu_Q = \mu_{Q_i}$, $\mu_u = \mu_{u_i}$, and $\mu_d = \mu_{d_i}$.
The Yukawa interactions for leptons yield
\begin{align}
	0 = \mu_{L_i} - \mu_H - \mu_{e_i}.
	\label{eq:y_l}
\end{align}
Here we have neglected any lepton flavour violation.\footnote{
In other words, for simplicity we are assuming that the right-handed neutrinos responsible for the neutrino mass generation are either sufficiently heavy or sufficiently weakly coupled to render lepton flavour violation negligible.
} In this approximation, the chemical potentials for the leptons can also depend on the generation.

\vspace{1em}\paragraph*{Conserved quantities.}
Within the SM, we have two conserved quantities associated with $B - L$ and U$(1)_Y$.
Neglecting lepton flavour violation, $B-L$  is moreover separately conserved for each generation, \textit{i.e.,} $B/3 - L_i$.
Thus, in total, we have four conserved quantities.
We are interested in the case with no initial asymmetries for these quantities, leading to the following four constraints:
\begin{align}
	0 &=  3 \left( \mu_{Q} + 2 \mu_{u} - \mu_{d} \right) - \sum_i \left( \mu_{L_i} + \mu_{e_i} \right) +  2\mu_H\,,
	\label{eq:neutral}
\end{align}
for U$(1)_Y$ and
\begin{align}
	0 &= 2 \mu_{Q} + \mu_{u} + \mu_{d} - 2 \mu_{L_i} - \mu_{e_i}\,,
	\label{eq:b-l}
\end{align}
for $B / 3 - L_i$.

\vspace{1em}\paragraph*{A bottleneck process.}
The final baryon asymmetry is determined by the competition between the complete thermalization and the chiral plasma instability, which we will discuss in the next subsection.
Instead of solving the equations for all the chemical potentials dynamically, we can reduce the degrees of freedom by just looking at the chiral plasma instability and a bottleneck process for the reshuffling of the chemical potentials.
Among them, the Yukawa interaction for the first generation lepton is the slowest process, which becomes efficient only for temperatures $\hat T \lesssim 10^5$\,GeV~\cite{Campbell:1992jd,Cline:1993bd,Bodeker:2019ajh}.
Above this temperature, the right-handed electron is decoupled from the others and hence the reshuffling does not changes the value of its chemical potential.

Assuming that all the interactions are efficient except for the Yukawa interaction of the first generation lepton, we can express all the chemical potentials in terms of $\mu_{e_1}$, the chemical potential of the right-handed electron.
One can verify this immediately by noting that  Eqs.~\eqref{eq:sph} -  \eqref{eq:b-l} then result in $9$ independent constraints for 10 degrees of freedom.\footnote{Recall that we have taken the same chemical potentials for quarks of different generations, \textit{i.e.}, $\mu_Q = \mu_{Q_i}$, $\mu_u = \mu_{u_i}$, and $\mu_{d} = \mu_{d_i}$. Also note that the condition for the strong sphaleron is redundant with Eq.~\eqref{eq:y_q}.
}
The corresponding relations are derived in Appendix~\ref{sec:chemical}.
Therefore we focus on the evolution of the chemical potential for the right-handed electron, which is governed by~\cite{Joyce:1997uy,Kamada:2016eeb,Kamada:2016cnb}
\begin{align}
	\frac{ \partial \mu_{e_1}}{\partial \eta} = - \frac{1}{T^2/3}\frac{\partial q_{\text{CS},Y}}{\partial \eta} - \Gamma_{Y_e} \frac{711}{481} \mu_{e_1}\,.
	\label{eq:anomaly_w_Ye}
\end{align}
Here we have assumed that $\mu_{e_1} / T \ll 1$, keeping only the leading order term.
Note that the last term comes from the Yukawa interaction, $Y_{e,11} e_1^\dag H^\dag L_1 + \text{H.c.}$, which leads to a term proportional to $\mu_{L_1} - \mu_H - \mu_{e_1}$. Using the relation \eqref{eq:summary}, one finds $\mu_{L_1} - \mu_H - \mu_{e_1} = -711\mu_{e_1}/481$. For temperatures $\hat T \gtrsim 10^5$, the contribution from the last term is strongly suppressed because the Yukawa interaction is not in equilibrium.

\vspace{1em}
\paragraph*{Chiral magnetohydrodynamics.}

The evolution of the large-scale hyper gauge field which we are interested in is well described by the MHD approximation \cite{Durrer:2013pga},
\begin{align}
	0 = \frac{\partial \bm{B}_Y}{\partial \eta} + \bm{\nabla} \times \bm{E}_Y\,, \quad
	0 = \bm{\nabla} \times \bm{B}_Y - \bm{J}_Y\,, \label{eq:mhd_maxwell}
\end{align}
where the induced current is estimated by the generalized Ohm's law,
\begin{align}
	\bm{J}_Y = \sigma_Y \left( \bm{E}_Y + \bm{v} \times \bm{B}_Y \right) + \frac{2 \alpha_Y}{\pi} \mu_{Y,5} \bm{B}_Y\,,
	\label{eq:ohm}
\end{align}
for $\mu_{Y,5} / T \ll \alpha_Y$.\footnote{
See Ref.~\cite{Figueroa:2017hun} and therein for the classification of the regimes.
}
Here $\bm{v}$ represents the fluid velocity and $\sigma_Y \simeq  c_\sigma T/(\alpha_Y \ln(\alpha_Y^{-1})) \simeq 10^2  \,T$ with $c_\sigma \simeq 4.5$~\cite{Baym:1997gq,Arnold:2000dr} denotes the plasma conductivity. 
Note that all the quantities given here are comoving ones, and the corresponding physical quantities are given by \textit{e.g.}, $\hat \sigma_Y = \sigma_Y / a$ and $\hat \mu_{Y,5} = \mu_{Y,5} / a$.
The part of the induced current proportional to the hyper magnetic field is provided by the chiral magnetic effect with~\cite{Vilenkin:1980fu,Joyce:1997uy, Alekseev:1998ds, Son:2004tq, Fukushima:2008xe},
\begin{align}
	\mu_{Y,5} = \sum_\alpha \epsilon_\alpha N_\alpha Q_{Y, \alpha}^2 \mu_\alpha\,.
	\label{eq:mu5def}
\end{align}
Combining Eqs.~\eqref{eq:mhd_maxwell} and \eqref{eq:ohm}, we obtain
\begin{align}
	\frac{\partial \bm{B}_Y}{\partial \eta} = \frac{\bm{\nabla}^2}{\sigma_Y} 
	 \bm{B}_Y
	+ \bm{\nabla} \times \left( \bm{v} \times \bm{B}_Y \right)
	+ \frac{2 \alpha_Y}{\pi} \frac{\mu_{Y,5}}{ \sigma_Y} \bm{\nabla} \times \bm{B}_Y\,.
	\label{eq:B_field}
\end{align}

On top of this, the velocity field obeys the Navier-Stokes equation
\begin{align}
	 \frac{\partial}{\partial \eta} \bm{v} + \bm{v} \cdot \bm{\nabla} \bm{v}  = \nu \bm{\nabla}^2 \bm{v} + \frac{1}{\rho + P}\left( - \frac{1}{2} \bm{\nabla} \bm{B}_Y^2 + (\bm{B}_Y \cdot \bm{\nabla}) \bm{B}_Y \right) 
	\label{eq:velocity}
\end{align}
for an incompressible fluid ($\bm{\bm{\nabla} \cdot \bm{v}} = 0$).
Here $\rho$ and $P$ denote the energy density and pressure of the plasma, respectively, and $\nu$ represents the shear viscosity normalized by $\rho + P$, \textit{i.e.}, the kinetic viscosity, which can be estimated as $\nu = c_\nu /(\alpha_Y^2 \ln (\alpha_Y^{-1}) T) \simeq 10 / T$ with $c_\nu \simeq 0.01$~\cite{Durrer:2013pga,Arnold:2000dr}. 
These equations for the hyper magnetic field \eqref{eq:B_field}, velocity field \eqref{eq:velocity}, and chiral asymmetry \eqref{eq:anomaly_w_Ye} govern the chiral MHD.

In particular, we are interested in how the hyper magnetic field behaves after reheating. For this purpose, we need to understand the relative importance of each term in Eq.~\eqref{eq:B_field}.
Here the first term accounts for the decay of the hyper magnetic field driven by diffusion processes at small scales. The second term is crucial to describe inverse cascade processes in the fluid, which will transfer energy from smaller to larger scales. The third term encodes the effects of a non-zero chemical potential and will be crucial in the description of the chiral plasma instability. 
The characteristic time scales of these terms define qualitatively different regimes in the evolution of the helical hyper magnetic field as we will outline in the following.

\vspace{1em}
\paragraph*{Inverse cascade.}
Here we provide a condition to avoid magnetic diffusion so that the hyper magnetic helicity survives until the EW phase transition, which is an essential condition for our baryogenesis scenario to work.
Deriving a robust criterion to avoid magnetic diffusion requires a numerical simulation of MHD,
which is however beyond the scope of this paper.
In the following, we provide an order of magnitude estimation.

Let us first neglect the last term in Eq.~\eqref{eq:B_field} which leads to the chiral plasma instability. This approximation is justified at the early stage of the dynamics as we will discuss later.
Right after inflation, the correlation length of the hyper magnetic field is given by the comoving horizon scale,
\begin{align}
	L_\text{rh} \sim \frac{1}{a_\text{rh}H_\text{rh}}\,.
\end{align}
One might think that diffusion will erase the primordial hyper magnetic field.
In fact, we can easily see that, if the second term in Eq.~\eqref{eq:B_field} can also be neglected, the hyper magnetic field is dissipated away at
\begin{align}
	\eta_\text{diff} \sim \sigma_Y L_\text{rh}^2
	\quad \leftrightarrow \quad
	\hat T_\text{diff} \sim \frac{\alpha_Y \ln \alpha_Y^{-1}}{c_\sigma} H_\text{rh}\,.
\end{align}

To prevent this from happening, the second term in Eq.~\eqref{eq:B_field} has to be effective compared to the first term.
This term describes a coupling between the magnetic field and the velocity field, and essentially drives the magnetic field from smaller to larger scales. This process is called the inverse cascade.
As a result, if the inverse cascade occurs, magnetic diffusion becomes less efficient.
The second term in Eq.~\eqref{eq:B_field} comes into play at the eddy turnover scale:
\begin{align}
	\eta_\text{t} \sim \frac{L_\text{rh}}{v_\text{rh}}\quad \leftrightarrow \quad
	\hat T_\text{t} \sim v_\text{rh} \, \hat T_\text{rh}\,.
\end{align}
Here $v_\text{rh}$ is a typical amplitude of the velocity field at the scale $L_\text{rh}$.
We require that the eddy turnover happens before the magnetic diffusion becomes efficient, which is equivalent to having an initial magnetic Reynolds number larger than unity
(see \cite{Durrer:2013pga} and references therein), \textit{i.e.},
\begin{align}
	\eta_\text{t} < \eta_\text{diff} \quad \leftrightarrow \quad
	1 < R_{m, \text{rh}} \simeq \sigma_Y L_\text{rh} v_\text{rh}\,.
	\label{eq:magnetic_R1}
\end{align}

The amplitude of the velocity field can be estimated by balancing the hyper magnetic field and the velocity field in Eq.~\eqref{eq:velocity}.
If the initial kinetic Reynolds number is greater than unity,
\begin{align}
	1 < R_{e, \text{rh}} \simeq \frac{L_\text{rh} v_\text{rh}}{\nu}\,,
	\label{eq:kin_reynolds}
\end{align}
the induction term, $\bm{v} \cdot \bm{\nabla} \bm{v}$, dominates over the diffusion term, $\nu \bm{\nabla}^2 \bm{v}$. Balancing the induction term with the source term from the hyper magnetic field in Eq.~\eqref{eq:velocity}, we then expect an equipartition of the energy densities in the velocity field and the magnetic field, $\rho \, v^2 \sim B_Y^2$. This is supported by numerical simulations~\cite{Kahniashvili:2012uj,Banerjee:2004df}.\footnote{
Note, however, that for a maximally helical field, total equipartition is not reached and one instead finds $\rho \, v^2 \sim 0.1 \,B_Y^2$~\cite{Banerjee:2004df}.}
Note that, for the SM plasma at high temperatures, we always get $R_m \gg R_e$.
If $R_m \gg R_e > 1$, the system enters the turbulent regime, and each field then obeys a specific scaling relation such that the total helicity is approximately conserved.
This is derived in Appendix~\ref{sec:reynolds} [Eq.~\eqref{eq:scaling_turb}].
Inserting the equipartition relation $v^2 \sim B_Y^2 / \rho$ into Eq.~\eqref{eq:kin_reynolds}, one may put a lower bound on the initial hyper magnetic field to develop turbulence.

However, we find that Eq.~\eqref{eq:kin_reynolds} does not hold in most of the parameter space of our interest.
In this case, the amplitude of the velocity field may be estimated by balancing the diffusion term, $\nu \bm{\nabla}^2 \bm{v}$, and the source term from the hyper magnetic field:
\begin{align}
	v \sim \frac{L B_Y^2}{\rho \nu} \quad \leftrightarrow \quad \rho \, v^2 \sim R_e B_Y^2\,.
	\label{eq:velocity_smallRe}
\end{align}
It is clear that equipartition is not reached for $R_e < 1$.
Inserting this estimate into Eq.~\eqref{eq:magnetic_R1}, we can put a rough lower bound on the initial hyper magnetic field to avoid magnetic diffusion:
\begin{align}
	1 < R_{m, \text{rh}} \sim \frac{c_\sigma \alpha_Y}{c_\nu} \left( \frac{M_\ast}{\hat T_\text{rh}} \right)^2 \left( \frac{\hat B_Y^2}{\hat \rho_\text{inf}} \right)_\text{rh}\,.
	\label{eq:magnetic_R}
\end{align}
Here $\hat \rho_\text{inf}$ is the energy density during inflation and the effective number of degrees of freedom for relativistic particles, $g_\ast$, is absorbed into $M_\ast \equiv [90 / (\pi^2 g_\ast)]^{1/2} \Mpl$  with $g_* = 427/4$ for the SM above the EW phase transition.

Once the condition \eqref{eq:magnetic_R} is fulfilled, the relevant quantities %may obey 
feature an equilibrium solution which obeys
the following scaling:
\begin{align}
B_{Y,\eta} \propto \eta^{-1/2}\,, \quad  L_\eta \propto \eta\,, \quad  v_\eta \sim \text{const.}
\label{eq:scaling_viscous}
\end{align}
This is derived in Appendix~\ref{sec:reynolds} [Eq.~\eqref{eq:scaling_vis}].
These scaling relations are a consequence of helicity conservation and our estimate for the velocity field in Eq.~\eqref{eq:velocity_smallRe}. They were previously derived in Ref.~\cite{Banerjee:2004df} together with scaling relations for the case where the dissipation of the velocity field is due to free streaming of particles and not diffusion. Only the scaling relations for the latter case were compared with a numerical simulation in Ref.~\cite{Banerjee:2004df} but were confirmed with very good accuracy. We therefore expect that our procedure of estimating the velocity field in the regime $R_m > 1 > R_e$ gives a reasonable approximation also for our case where dissipation is due to diffusion. Nevertheless, a numerical simulation to verify this estimate is clearly warranted but is beyond the scope of this work.

Eq.~\eqref{eq:magnetic_R} [see also Eq.~\eqref{eq:magnetic_R1}] is sufficient to prevent magnetic diffusion at all times, since
\begin{align}
	\frac{\eta}{\sigma_Y L_\eta^2}  \lesssim \frac{\eta_\text{t}}{\sigma_Y L_\text{rh}^2} \sim \frac{1}{\sigma_Y L_\text{rh} v_\text{rh}} \ll 1\,.
\end{align}
Also, both the magnetic and kinetic Reynolds numbers grow with time, \textit{i.e.}, $R_m, R_e \propto \eta$, once the system falls into the scaling solution~\eqref{eq:scaling_viscous}.
Hence, at some point, the system may flow into the turbulent regime.
Now it is clear that this is a delicate issue of the initial conditions, and hence we would like to emphasize again that Eq.~\eqref{eq:magnetic_R} should not be regarded as a sharp cutoff, rather just as a rough indicator.

Alternatively, one may put a more conservative bound by not referring to the dynamics of the velocity field.
Assuming that the velocity field is generated from the hyper magnetic field, we expect that its energy is at most the energy density carried by the hyper magnetic field, \textit{i.e.}, $\rho\, v^2 \lesssim B_Y^2$.
Hence, the magnetic Reynolds number is bounded from above, $R_{m, \text{rh}} \lesssim R_{m, \text{rh}}^\text{max} = \sigma_Y L_\text{rh} (\hat B_Y^2 / \hat \rho_\text{inf})^{1/2}$.
To avoid magnetic diffusion, we at least need $1 < R_{m, \text{rh}}^\text{max}$ which gives
\begin{align}
	1 < R_{m, \text{rh}}^\text{max} \sim 
	\frac{c_\sigma}{\alpha_Y \ln \alpha_Y^{-1}} \left( \frac{M_\ast}{\hat T_\text{rh}} \right)
	\left(\frac{\hat B_Y^2}{\hat \rho_\text{inf}} \right)^\frac{1}{2}_\text{rh} \,.
	\label{eq:Rm_max}
\end{align}

\vspace{1em}
\paragraph*{Chiral plasma instability.}
Now we are ready to discuss how the chiral plasma instability arising from the last term in Eq.~\eqref{eq:B_field} can change the picture.
The value of $\mu_{Y,5}$ right after inflation can be expressed by $q_{\text{CS},Y}$ since the Yukawa couplings and sphalerons are not efficient during inflation [see Eqs.~\eqref{eq:charges anomaly}, \eqref{eq:rel_chempot_charge} and \eqref{eq:mu5def}]:
\begin{align}
	\mu_{Y,5}^\text{rh} = \frac{95}{54} \mu_{B+L}^\text{rh} = - \frac{95}{18} \frac{q_{\text{CS},Y}^\text{rh} }{T^2/3}\,.
	\label{eq:initial}
\end{align}
Although the reshuffling of the chemical potentials in the radiation dominated era will modify the numerical factor in this equation, $\mu_{Y,5}$ will remain proportional to $q_{\text{CS},Y}$ unless all the interactions enter thermal equilibrium. 
For instance, if all the interactions become efficient except for the electron Yukawa, we find [see Eq.~\eqref{eq:summary} in Appendix~\ref{sec:chemical}]
\begin{align}
	\mu_{Y,5} = \frac{79}{44} \mu_{B+L} =  \frac{711}{481} \mu_{e_1} =  - \frac{711}{481} \frac{q_{\text{CS},Y}}{T^2/3}\,.
	\label{eq:mu_5_ye}
\end{align}
Prior to the chiral plasma instability, the helical charge retains its initial value, $h_\text{rh} = ( \pi / \alpha_Y)\,  q_{\text{CS},Y}^\text{rh} $, even after the reshuffling, because of the approximate conservation of helicity for $R_m > 1$.

However, once the last term in Eq.~\eqref{eq:B_field} becomes relevant, the absolute value of the helical charge, $\abs{h}$, starts to decrease.
To see this clearly, it is convenient to write down the evolution equation for the helical charge.
Using Eqs.~\eqref{eq:mhd_maxwell} and \eqref{eq:ohm}, one may express $\bm{E}_Y$ in terms of $\bm{B}_Y$ and $\bm{v}$:
\begin{align}
	\bm{E}_Y = \frac{1}{\sigma_Y} \left( \bm{\nabla} \times \bm{B}_Y - \frac{2 \alpha_Y}{\pi} \mu_{Y,5} \bm{B}_Y \right) - \bm{v} \times \bm{B}_Y.
	\label{eq:hyper_E}
\end{align}
Inserting this into Eq.~\eqref{eq:cs-evolution} with $q_{\text{CS}, Y} = (\alpha_Y / \pi) h$ yields\footnote{Note that the helicity does not depend explicitly on the velocity field because $\bm{B}_Y \cdot (\bm{v} \times \bm{B}_Y) = 0$, indicating that the velocity-dependent term in Eqs.~\eqref{eq:B_field} and \eqref{eq:hyper_E} conserves helicity. A dependence on the velocity field however still arises indirectly because it drives the typical length scale of the hyper gauge field towards the infrared. See also Eqs.~\eqref{eq:scaling_turb} and \eqref{eq:scaling_vis}.
}

\begin{align}
	\frac{\partial}{\partial \eta} h = \int \frac{\dd^3 x}{\text{vol}\, (\mathbb{R}^3)} \, \left(
		2 \bm{B}_Y \cdot \frac{\bm{\nabla}^2}{\sigma_Y} \bm{A}_Y + \frac{4 \alpha_Y}{\pi} \frac{\mu_{Y,5}}{\sigma_Y} \bm{B}_Y^2
	\right)\,.
	\label{eq:helical}
\end{align}
For our purpose, it is convenient to perform a Fourier transformation of Eq.~\eqref{eq:helical} (see Appendix \ref{sec:cpi} for details),
\begin{align}
	\frac{\partial}{\partial \eta} h_k (\eta) = 
	\frac{2 k}{\sigma_Y} \left( \frac{k_\text{CPI}}{r_k(\eta)} - k \right) h_k (\eta)\,,
	\label{eq:FTqcs}
\end{align}
where the degree of helicity is encoded in
\begin{align}
	r_k(\eta) := \frac{k \, h_k (\eta) / 2}{\rho_{B,k} (\eta)}\,,
\end{align}
with $\rho_{B,k}$ being the energy density of the hyper magnetic field mode with momentum $k$, $-1 \leq r_k \leq 1$, and $|r_k| = 1 $ denoting a maximally helical configuration.
Eq.~\eqref{eq:FTqcs} defines the characteristic scale for the chiral plasma instability~\cite{Akamatsu:2013pjd,Yamamoto:2016xtu}:%
\footnote{This can also be understood as follows \cite{Joyce:1997uy}: The charge density of the right-handed electrons is $\sim \mu_{e_1} T^2$, while the corresponding energy density is $\sim\mu_{e_1}^2 T^2$. As follows from Eq.~\eqref{eq:anomalous}, this charge density can be converted into a configuration of the hyper gauge field with $\alpha_Y A_Y^2 / L \sim \mu_{e_1} T^2 $, where $L$ is a characteristic length scale. This gauge field configuration has an energy density ${\sim A_Y^2/L^2}$ and for $L \gtrsim L_{\rm CPI}\equiv (\alpha_Y \mu_{e_1})^{-1} = k_{\rm CPI}^{-1}$ it becomes energetically favorable compared to the charge density of the right-handed electrons. This reproduces Eq.~\eqref{k_CPI} (up to numerical factors). }
\begin{align}
\label{k_CPI}
	k_\text{CPI} := \frac{2 \alpha_Y}{\pi} \mu_{Y,5}\,.
\end{align}
For a maximally helical mode with $|r_k| = 1$, $|k_\text{CPI}|$ gives the upper boundary of the $k$-range which is affected by the chiral plasma instability.
From Eqs.~\eqref{eq:initial} and \eqref{eq:mu_5_ye}, we see that $\mu_{Y,5} (\eta) \lessgtr 0$ for $q_{\text{CS},Y} (\eta) = (\alpha_Y / \pi) h (\eta) \gtrless 0$. One then finds growing modes $h_k (\eta)$ of the helicity which have opposite sign compared to the total helicity $h (\eta)$.
As a result, the total integrated helical charge decreases eventually.
The fastest growing mode is $k \sim k_\text{CPI}/2$ and the time scale of its growth can be estimated as 
$\eta_\text{CPI}  \sim 2 \sigma_Y/k_\text{CPI}^2  = \pi^2 \sigma_Y/(2 \alpha_Y^2 \mu_{Y,5}^2)$. The corresponding temperature is given by~\cite{Kamada:2018tcs}
\begin{align}
	\hat T_\text{CPI} &\sim 1.5 \times 10^5 \GeV \; \times \frac{\ln(\alpha_Y^{-1})}{c_\sigma} \left( \frac{100}{g_\ast} \right)^\frac{1}{2} \left( \frac{\alpha_Y}{0.01} \right)^3 \left( \frac{\mu_{Y,5} / T}{10^{-3}}\right)^2 \nonumber \\
	&\sim 1.5 \times 10^5 \GeV\; \times \frac{c_\mu^2 \ln(\alpha_Y^{-1})}{c_\sigma } \left( \frac{g_\ast}{100} \right)
	\left( \frac{\alpha_Y}{0.01} \right)^{5} \left( \frac{H_\text{rh}}{10^{14} \GeV} \right)^3 \nonumber \\
	& \qquad \qquad \qquad \; \; \; \; \; \; \; \;  \times\left( \frac{\vev{\hat{\bm{E}}_Y\cdot \hat{\bm{B}}_Y}_\text{rh} / H_\text{rh}^4}{10^5} \right)^2\,, \label{eq:TCPI}
\end{align}
where we have parametrized the time-dependent relation between $\mu_{Y,5}$ and $q_{\text{CS},Y}$ due to the reshuffling of the chemical potentials by $\mu_{Y,5} = - c_\mu q_{\text{CS},Y}/(T^2/3)$.
For instance, one finds $c_\mu = 95/18$ initially [Eq.~\eqref{eq:initial}], $c_\mu = 711/481$ [Eq.~\eqref{eq:mu_5_ye}] when all the interactions except for the electron Yukawa are efficient, and $c_\mu = 0$ when all interactions enter thermal equilibrium.

Let us first consider the case where the chiral plasma instability occurs before the electron Yukawa becomes efficient: $\hat T_\text{CPI} \gg \hat T_{Y_{e,11}} \sim 10^5 \GeV$.
In this regime, we have an approximate conservation law for $\mu_{e_1} + q_{\text{CS},Y}/(T^2/3)$ as can be seen from Eq.~\eqref{eq:anomaly_w_Ye}.
Thus the decay of $q_{\text{CS},Y}$ via the chiral plasma instability leads to the decay of $\mu_{e_1}$.
Since $\mu_{e_1} + q_{\text{CS},Y}/(T^2/3) = 0$ holds even after the reshuffling of chemical potentials, this process could erase both $\mu_{e_1}$ and $q_{\text{CS},Y}$.
To avoid this situation, we conservatively require that the chiral plasma instability does not occur before the electron Yukawa comes into equilibrium: $\hat T_\text{CPI} \ll \hat T_{Y_{e,11}} \sim 10^5 \GeV$.
In this case, the electron Yukawa first erases the primordial asymmetry stored in $\mu_{e_1}$ as can be seen from Eq.~\eqref{eq:anomaly_w_Ye}.
As a result, we no longer expect the chiral plasma instability because $\mu_{Y,5} \propto \mu_{e_1}$ becomes negligible, and hence the Chern-Simons charge, $q_{\text{CS},Y}$, survives.
As we see in the next section, this surviving $q_{\text{CS},Y}$ regenerates a $B+L$ asymmetry at the EW phase transition.

Before concluding this section, let us emphasize that the MHD simulations supporting the scaling behavior~\eqref{eq:scaling_viscous} [see also~\eqref{eq:scaling_vis}] for $R_m > 1$ have not been performed {for the initial condition of our interest} including large chemical potentials as required by Eq.~\eqref{eq:initial}. See Refs.~\cite{Brandenburg:2017rcb, Schober:2017cdw, Schober:2018ojn} for studies of MHD turbulence with chiral asymmetries for different initial conditions. We therefore do not make any conclusive statements about the interplay of the velocity field and the chemical potentials in the regime where both the second and third term of Eq.~\eqref{eq:B_field} are important. We can, however, self-consistently require the chemical potentials to be inefficient while assuming that magnetic diffusion is suppressed by the velocity field of the plasma.
This is precisely the condition that we need to satisfy in order to ensure successful baryogenesis as we demonstrate in the next section.

%%%%%%%%%%%%%%%%%%%%%%%%%%%%%%%%%%%%%%%%%%%%%%%%%%%%%%%%%%%%%55

\begin{figure*}
\center
 \includegraphics[width = 0.95 \columnwidth]{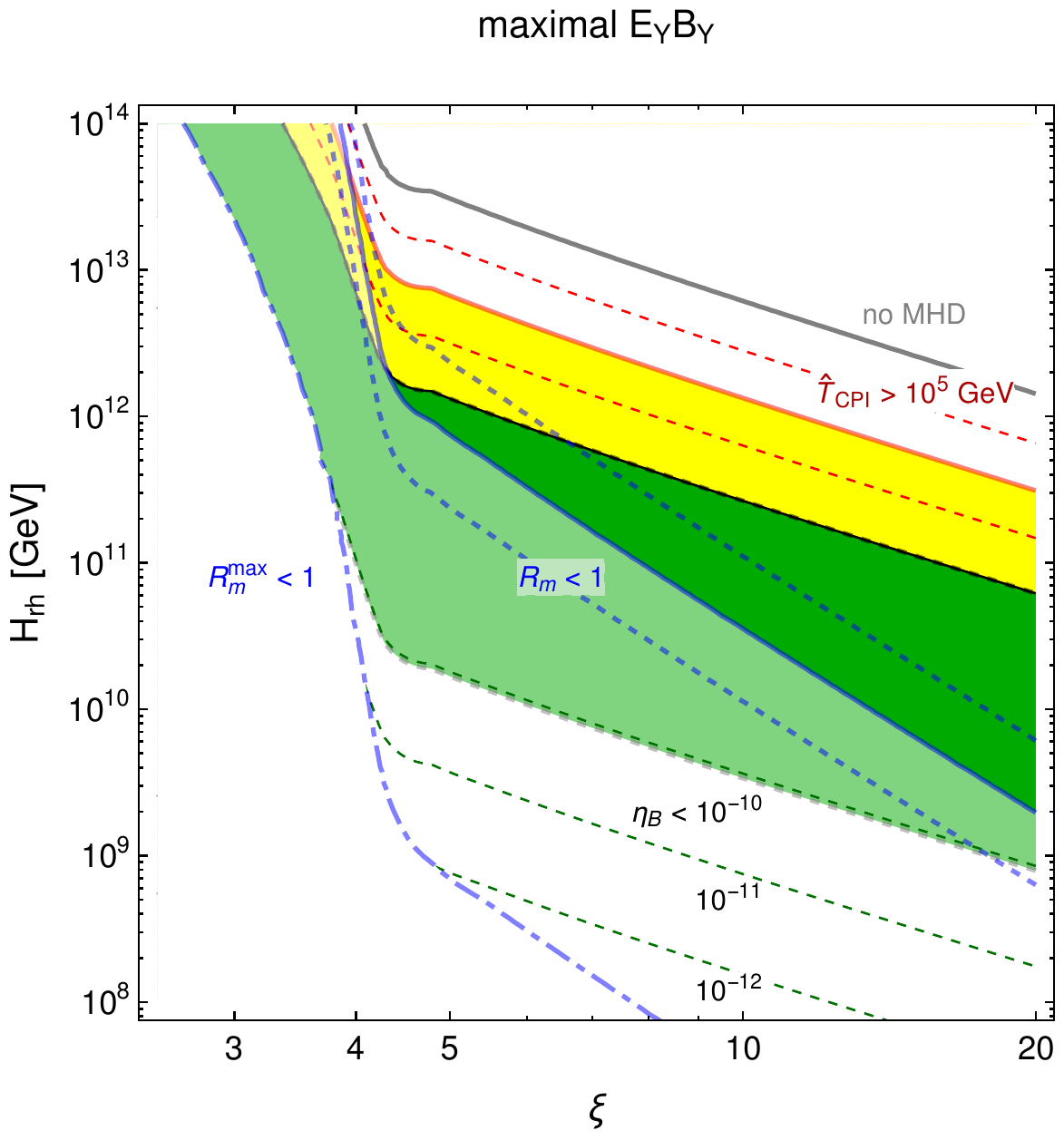} \hspace{0.1 \columnwidth}
 \includegraphics[width = 0.95 \columnwidth]{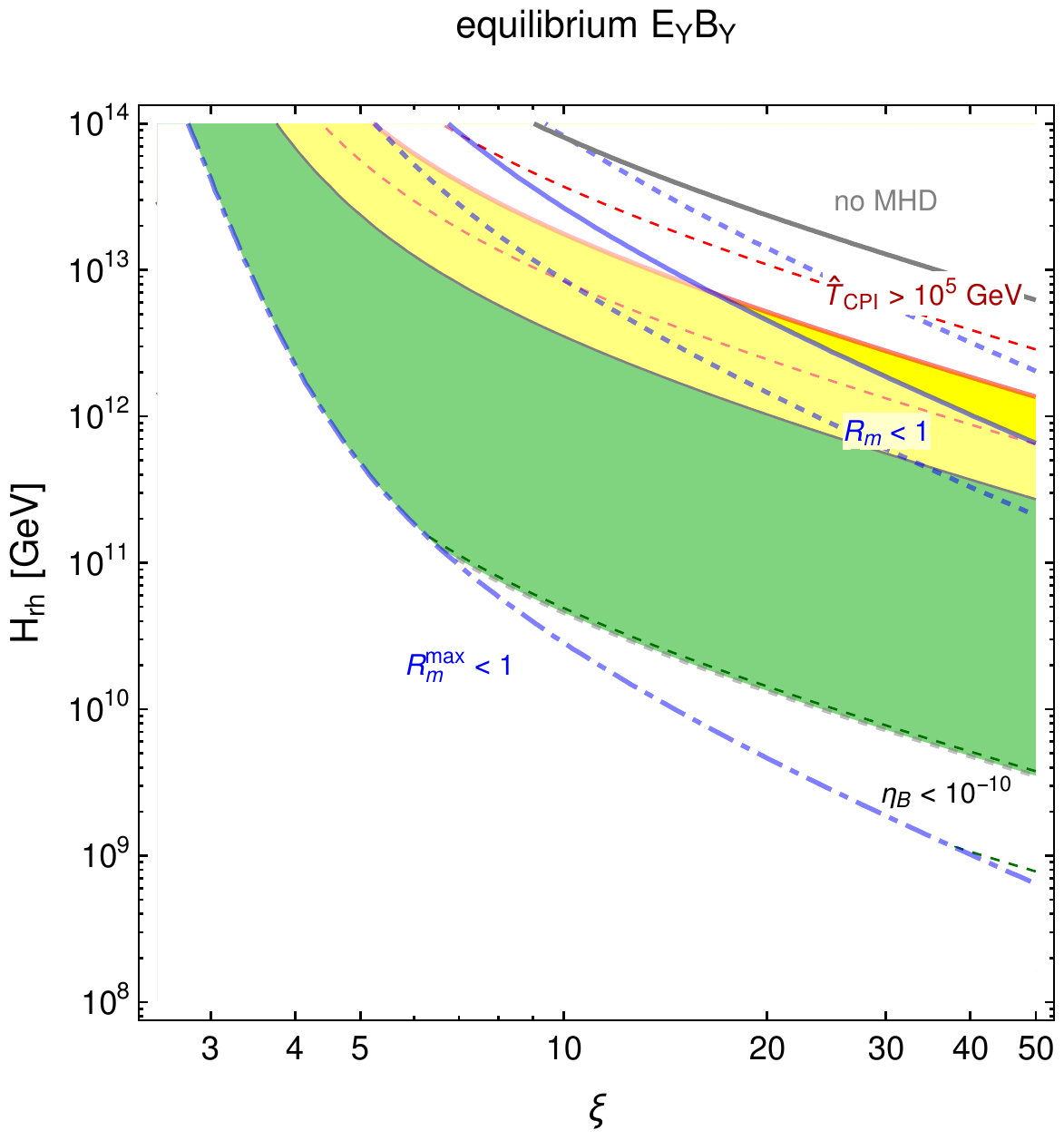}
  \caption{Baryon asymmetry of the Universe compatible with the observed value assuming the upper bound on $\hat E_Y \hat B_Y$ is saturated (left panel) or assuming the equilibrium value for $\hat E_Y \hat B_Y$ (right panel). The dark green bands indicate a baryon asymmetry in agreement with the observed value, $\eta_\text{obs} \simeq 9 \times 10^{-11}$, for $f(\theta_W, \hat T)$ between $f_\text{max}$ (dashed line) and $f_\text{min}$ (solid line). 
The dashed green lines show contours of $\eta_B$ for $f = f_\text{max}$.
The yellow region indicates an overproduction of the baryon asymmetry, $\eta_B > \eta_\text{obs}$, whereas all white regions correspond to a too small baryon asymmetry. In particular, 
in the region below the solid blue line magnetic diffusion is likely to be efficient,
corresponding to a small magnetic Reynolds number. For reference, the dashed blue lines indicate $R_m = \{1/3, 3\}$, respectively, and the dashed-dotted blue line indicates  our most conservative estimate of the magnetic Reynolds number.
The region above the red line is excluded because the plasma instability occurs before the electron Yukawa coupling becomes relevant. For reference, the dashed red lines correspond to $\hat T_\text{CPI} = \{10^4, 10^6\}$~GeV, respectively. 
The gray line indicates the limitation of the MHD approximation. We recall that for $\xi \lesssim 4$ (left panel) and $\xi \lesssim 9$ (right panel) significant uncertainties might arise from the (model-dependent) reheating dynamics.}
 \label{fig:etaB}
\end{figure*}

\section{Baryogenesis at the EW phase transition \label{sec:baryogenesis}} 

\vspace{1em}
\paragraph*{Regeneration of the baryon asymmetry.}
Requiring that the inverse cascade prevents efficient diffusion of the hyper magnetic field and that the plasma instability does not occur before the electron Yukawa interactions reach equilibrium ($\hat T_\text{CPI} \ll 10^5$~GeV) ensures that at the EW phase transition ($\hat T_\text{EW} \sim 100$~GeV) all chemical potentials have been (approximately) erased whereas the comoving helicity stored in the hyper magnetic field has (approximately) conserved the value present at reheating, given by Eq.~\eqref{eq:qcs0}. 
At the EW crossover, the hyper magnetic field is smoothly converted into the magnetic field of $U(1)_\text{em}$. 
This conversion is controlled by the Weinberg angle, defined as the angle of the SO(2) rotation that diagonalizes the mass matrix of $W^3_\mu$ and the hyper gauge boson.
In the zero-temperature limit, the Weinberg angle changes discontinuously from $\tan\theta_W = 0$ in the symmetric phase to $\tan\theta_W = g_Y/g_W$ in the broken phase with $g_W$ and $g_Y$ being the SU$(2)_W$ and U$(1)_Y$ gauge couplings of the SM, respectively. Taking into account finite temperature corrections to the gauge boson masses, the evolution of the Weinberg angle becomes smooth, albeit subject to significant uncertainties~\cite{Kajantie:1996qd,DOnofrio:2015gop}.
In Ref.~\cite{Kamada:2016cnb}, it was found that this effect dominates the production of the baryon/lepton asymmetry.

To see this, one may focus on the dynamics after the EW phase transition, in particular when the EW sphaleron becomes the slowest process.
In this regime, we can solve the equations for the chemical potentials by assuming that all the SM interactions except for the EW sphaleron are in chemical equilibrium.
The resulting solution is derived in Appendix \ref{sec:chemical} [see Eq.~\eqref{eq:summary2}]. In order to derive an evolution equation for the baryonic charge in this regime, we consider the chiral anomaly equation for baryon number, 
\begin{align}
	\partial_\mu J^\mu_B = 3 \left( \frac{g_W^2}{32 \pi^2} W^{a \mu\nu} \tilde W^a_{\mu\nu} - \frac{g_Y^2}{32\pi^2} Y^{\mu\nu} \tilde Y_{\mu\nu} \right)\,.
\end{align}
Taking into account contributions to the right-hand side from the EW sphaleron and the helical hyper gauge fields, this gives
\begin{align}
	\partial_\eta q_B = - \frac{111}{34} T \gamma_{W,\text{sph}} q_B + \frac{3}{2} \left( g_W^2 + g_Y^2 \right) \sin (2 \theta_W) (\partial_\eta \theta_W) \frac{h}{8 \pi^2}\,. \nonumber
\end{align}
Here $\gamma_{W, \text{sph}} \simeq 3.3 \times 10^{-16}$~\cite{DOnofrio:2014rug} represents the dimensionless transport coefficient for the EW sphaleron.%
\footnote{As pointed out in \cite{Comelli:1999gt}, the energy of the EW sphaleron decreases in the presence of a strong hyper magnetic field. We have checked that the hyper magnetic field in our case is never strong enough to have a noticeable effect.}
The corresponding term in the evolution equation tends to reduce the baryon asymmetry.
Furthermore, the time evolution of $\theta_W$ parametrizes the smooth transition of the Weinberg angle at the EW crossover. The corresponding term in the evolution equation reflects the fact that the magnetic field of $U(1)_\text{em}$, into which the hyper magnetic field is converted, no longer contributes to the chiral anomaly of baryon number. It therefore acts as a source term for the baryon asymmetry~\cite{Kamada:2016cnb}.
Note that the first term on the right-hand side can be obtained by inserting the solution \eqref{eq:summary2} into the sphaleron reaction rate, resulting in a term proportional to $\frac{3}{2} \sum_i \left( 3 \mu_{U} + 3 \mu_{D} + \mu_{\nu_i} + \mu_{E_i} \right) = 111 \mu_B / 34$. See Appendix \ref{sec:chemical} for details.

One may find an attractor solution to the evolution equation for the baryonic charge by requiring that $\partial_\eta q_B = 0$, \textit{i.e.}, a dynamical equilibrium between the washout and production terms.
After the completion of the EW phase transition, $\gamma_{W,\text{sph}}$ becomes exponentially small and $\partial_\eta \theta_W$ goes to zero and the baryon asymmetry becomes frozen. The resulting asymmetry can be
estimated as,
\begin{align}
 \eta_B = \frac{q_B}{s} \simeq \frac{17}{37} \left[ (g_W^2 + g_Y^2) \frac{f(\theta_W,\hat T) \, {\cal S}}{\gamma_{W,\text{sph}}} \right]_{T = 135~\text{GeV}} \,,
 \label{eq:etaB}
\end{align}
where the temperature at which the expression is evaluated is determined from numerical simulations~\cite{Kamada:2016cnb}.
Furthermore, $ s = (2 \pi^2/45) \, g_*  T^3$ denotes the entropy of the SM thermal bath, 
$g_W \simeq 0.64$ and $g_Y \simeq 0.35$ are evaluated at the electroweak scale, $f(\theta_W,\hat T) = -\hat T \dd\theta_W/\dd\hat T \sin(2 \theta_W)$ encodes the change in the Weinberg angle $\theta_W$ around the EW phase transition, and ${\cal S} = H/(\hat s \hat T) h/(8 \pi^2 a^3)$
encodes the Chern-Simons charge.
The evaluation of $f(\theta_W,\hat T)$ in the SM comes with significant uncertainties, following Ref.~\cite{Jimenez:2017cdr} (see also \cite{Kamada:2016cnb}) we consider values in the range $f_\text{min} \lesssim f(\theta_W,\hat T) \lesssim f_\text{max}$ with
\begin{align}
 f_\text{min} = 5.6 \times 10^{-4} \,, \quad f_\text{max} = 0.32 \,.
 \label{eq:f}
\end{align}
Using Eq.~\eqref{eq:qcs0} and helicity conservation after reheating, we find
\begin{align}
 {\cal S} = -\frac{5}{8 \pi^3 \sqrt{10 \, g_*} } \frac{\langle \hat{ \bm{E}}_Y \cdot \hat{\bm{B}}_Y \rangle_{\text{rh}}}{\hat T^2 \Mpl H_\text{rh}} \left( \frac{\hat T}{\hat T_\text{rh}} \right)^3 \,.
\end{align}
Now we see explicitly that a positive baryon asymmetry requires anti-parallel electric and magnetic fields,  $\hat{\bm E}_Y \cdot \hat{\bm B}_Y < 0$, which is obtained for $\dot \phi < 0$.%
\footnote{This is consistent with the discussion in~\cite{Jimenez:2017cdr} which finds $\dot \phi > 0$ but uses the opposite sign convention for the $\phi Y \tilde Y$ term in Eq.~\eqref{eq:setup_conf1}.}

\vspace{1em}
\paragraph*{Viable parameter space for baryogenesis.}
Assuming instantaneous reheating, 
the baryon asymmetry in Eq.~\eqref{eq:etaB} depends only on the energy scale of inflation, $H_\text{rh}$, and the parameter $\xi$ governing the amplitude of the primordial hyper gauge fields $\langle \hat E_Y \hat B_Y \rangle_\text{rh}$. In Fig.~\ref{fig:etaB}, we show the resulting baryon asymmetry in this plane, together with the conditions required to sustain the helicity stored in the hyper magnetic field until the EW phase transition. For the left panel, we assume that the hyper gauge fields generated during inflation saturate the upper bound, as depicted by the solid lines in Fig.~\ref{fig:EB}. For the right panel, we assume fast equilibration during inflation, \textit{i.e.}, that these hyper gauge fields are given by the equilibrium solution indicated by the dashed lines in Fig.~\ref{fig:EB}. Note that in this case, the parameter relevant for gauge field production is $\xi_\text{eff}$ defined in Eq.~(\ref{eq:xi_eff}), explaining the larger plot range in the right panel. In both cases, the dark green shaded area denotes the region where the observed value of the baryon asymmetry is reproduced, with the width of the band accounting for the uncertainty in the time-evolution of the Weinberg angle throughout the EW phase transition, as parametrized by Eq.~\eqref{eq:f}. In the region above the solid red line, the chiral plasma instability happens before the electron Yukawa becomes relevant, erasing both the helicity stored in the gauge fields and the chemical potentials of the fermions. 
Below the solid blue line the magnetic Reynolds number [Eq.~\eqref{eq:magnetic_R}] is too small, preventing the system from entering the turbulent or viscous regime.  In this case, diffusion will erase the helicity of the hyper magnetic field. 
We emphasize that this bound is an estimate only (see App.~\ref{sec:reynolds}) and should only be taken as an indication for the boundary of the regime in which magnetic diffusion is efficient.
For reference, the dashed-dotted blue contour shows our most conservative estimate of this boundary [Eq.~\eqref{eq:Rm_max}]. 
In summary, below the blue line and above the red line, the baryon asymmetry of the Universe is exponentially suppressed and essentially zero. It is intriguing to note that in the remaining band and within the theoretical uncertainties, the baryon asymmetry is enforced to be very close to the observed value.
To emphasize this point, the dashed green lines indicate contours of $\eta_B$ for $f = f_\text{max}$. Finally we note that the MHD approximation we employ is only valid below the solid gray line ($\mu_{Y,5}/T < \alpha_Y$), which however covers all the relevant parameter space.

%We point out that the kink in the contours at $\xi \simeq 15$ in the left panel of Fig.~\ref{fig:etaB} is due to the saturation of the energy conservation bound during inflation (see Fig.~\ref{fig:EB}). As discussed in Sec.~\ref{sec:axioninflation}, this indicates that in this regime the gauge field production will not saturate the upper bound derived from Eq.~\eqref{eq:curve}, and we should take the large $\xi$ region of this panel with a grain of salt. For the equilibrium solution of $\hat E_Y \hat B_Y$ this constraint is only reached at $\xi \sim 50$, explaining the larger plot range in the right panel.

Fig.~\ref{fig:etaB} has several important consequences. First, the baryon asymmetry is either completely erased (in which case there are no observers to wonder about it) or it is predicted to be in the range $6 \times 10^{-13} < \eta_B < 4 \times 10^{-7}$ (accounting for the allowed range of $f_\text{min} < f(\theta_W, \hat T) < f_\text{max}$). The observed value, $\eta_\text{obs} \simeq 9 \times 10^{-11}$ is perfectly consistent with this prediction.  The predicted range of $\eta_B$ would be narrowed down significantly (potentially to the point of falsifying this scenario) if the theoretical uncertainty in $f(\theta_W, \hat T)$ and/or the estimate of the time-scale of the chiral plasma instability could be reduced and/or the competition between diffusion and turbulence for $R_e < 1 < R_m$ were better understood and/or the non-linear dynamics of the fermion backreaction during inflation could be solved.
We hope that our work will trigger a more detailed investigations of these questions, in particular by means of dedicated MHD simulations. 
On the contrary, we stress that our predictions for the baryon asymmetry are only mildly dependent on the details of the underlying inflation model. In particular after imposing the constraints from magnetic diffusion and the plasma instability, the predicted value of the baryon asymmetry is 
only mildly sensitive to the value of $\xi$.

Second, we point out that the predicted baryon asymmetry increases for large $H_\text{rh}$ and $\xi$. This implies that for $\hat T \sim \hat T_\text{CPI}$ there is a competition between the overproduction of the baryon asymmetry due to a large helicity in the gauge fields (as indicated by the yellow band in Fig.~\ref{fig:etaB}) and the erasure time-scale of this helicity due to the chiral plasma instability. 
In this regime it becomes crucial to accurately model the competing dynamics of the electron Yukawa coupling, the magnetic diffusion and the chiral plasma instability. A full analysis taking all of these effects into account might re-open some parameter space around the solid red lines in Fig.~\ref{fig:etaB}. 
This applies in particular to the equilibrium solution for the hyper magnetic fields generated during inflation (right panel of Fig.~\ref{fig:etaB}), since here a region of baryon asymmetry overproduction is sandwiched between two suppression processes: the chiral plasma instability and diffusion. A detailed analysis of this question is however
beyond the scope of the present paper.

Third, we note that once the fermion backreaction becomes relevant at $\xi \gtrsim 4$, $\eta_B$ is essentially determined by the scale of inflation with only a very mild dependence on $\xi$.  In particular, successful baryogenesis  imposes a lower bound on the energy scale of inflation, $H_\text{rh} \gtrsim 10^{10}$~GeV.

%%%%%%%%%%%%%%%%%%%%%%%%%%%%%%%%%%%%%%%%%%%%%%%%%%%%%%%%%%%%%

\section{Late time magnetic fields and structure formation \label{sec:furtherobs}}

After the EW phase transition, the helical magnetic field will be diluted due to the adiabatic expansion of the Universe. Its remnant at late times will contribute to the intergalactic magnetic field, permeating the voids in our Universe. The observed suppression of secondary emission in the GeV range in the $\gamma$-ray spectrum of blazars may indicate a lower bound on this intergalactic magnetic field, whose origin is a subject of active research (see \textit{e.g.}, \cite{Turner:1987bw,Ratra:1991bn,Garretson:1992vt,Field:1998hi,Anber:2006xt,Durrer:2010mq,Ferreira:2013sqa,Ferreira:2014hma,Kobayashi:2014sga,Caprini:2014mja,Fujita:2016qab,Fujita:2019pmi} and references therein). In addition, CMB observations impose an upper bound on long range magnetic fields, see Ref.~\cite{Durrer:2013pga} for a review.

Contrary to the baryon asymmetry, the value of the present day magnetic field depends on the scaling laws in the regime with {$R_m > 1$}, see Eqs.~\eqref{eq:scaling_turb} and \eqref{eq:scaling_vis}. Eq.~\eqref{eq:scaling_turb} yields a slower decay of the hyper magnetic field and can hence be used to estimate an upper bound for the present day magnetic fields. Together with an
adiabatic expansion outside the turbulent region (see Ref.~\cite{Jimenez:2017cdr} for a detailed computation), we derive an upper bound for the strength 
of the present day magnetic field for a given hyper magnetic field strength at the end of inflation (with a correlation length of order $H_\text{rh}^{-1}$). Within the viable region for baryogenesis (corresponding to the green band in Fig.~\ref{fig:etaB}), we find magnetic field strengths $\hat B_{\text{em},0} \,{\lesssim} \, 10^{-17}\,$Gauss and correlation lengths $\gtrsim \, 0.1\,$parsec. These values are far below the observational upper bound on long range magnetic fields, but are also (by about three orders of magnitude) insufficient to provide an explanation for the blazar observations. We emphasize however that this is just a rough estimate which probably deserves a refinement by means of a dedicated MHD simulation accounting for the non-vanishing chemical potentials and the details of the viscous and turbulent regime.

The gauge field production during inflation provides an additional source for the primordial scalar and tensor power spectrum. These classical contributions to the power spectra have peculiar features~\cite{Barnaby:2011qe}. The contribution to the scalar power spectrum is highly non-Gaussian and thus strongly constrained at the scales probed by the CMB inhomogeneities. The contribution to the gravitational wave spectrum is chiral, distinguishing it from many astrophysical foregrounds. In single field inflation models, both the scalar and tensor power spectra can increase dramatically over the course of inflation, leading to interesting signatures at small scales. In the case of the scalar power spectrum, this leads to enhanced structure formation at small scales with the possibility of primordial black hole production~\cite{Linde:2012bt,Domcke:2017fix}. The chiral tensor power spectrum is an interesting target for ground- and space-based gravitational wave interferometers such as LIGO, LISA and the Einstein Telescope. The details of these signals depend on the scalar potential of the inflation model~\cite{Domcke:2016bkh} and may be strongly affected by the induced fermion current~\cite{Domcke:2018eki}. For these reasons we shall not discuss this phenomenology in any detail here, but simply point out that a range of upcoming experiments have the chance of collecting independent hints in favour of this baryogenesis scenario.

\section{Discussion and Conclusions \label{sec:conclusion}}

In this paper, we study baryogenesis in axion inflation through a Chern-Simons coupling between the inflaton and the U$(1)_Y$ of the SM gauge group.
During inflation, the non-vanishing velocity of the inflaton sources the dual production of helical hyper gauge fields and a $B+L$ asymmetry, as indicated by the anomaly equation.
Once the thermal plasma is generated after inflation,
the $B+L$ asymmetry tends to be erased via sphaleron and Yukawa interactions while the helical hyper gauge fields may survive if magnetic diffusion and the chiral plasma instability are not efficient.
Due to this hierarchy in the erasure of the $B+L$ asymmetry and the helical gauge fields, a sufficient amount of baryon asymmetry can remain after the sphaleron decoupling.

By taking into account (1) the backreaction from the induced current, (2) the efficiency of magnetic diffusion quantified by the magnetic Reynolds number, and (3) the chiral plasma instability,
we find a viable parameter region for baryogenesis in the $(\xi, H_\text{rh})$ plane (Fig.~\ref{fig:etaB}).
Although the latter two conditions constrain the viable parameter space from both sides, the resulting baryon asymmetry is consistent with the observed value within the theoretical uncertainties.
Outside of this region, the baryon asymmetry goes to zero, indicating (up to theoretical uncertainties) a definite prediction for the observed matter-antimatter asymmetry.
Interestingly, for $\xi \gtrsim 4$ the baryon asymmetry almost solely depends on the inflation scale, implying $H_\text{inf} \sim 10^{10}$--$10^{12}$\, GeV to reproduce the observed value.

Ultimately, our prediction {for the baryon asymmetry} should be a line instead of a band in the $(\xi, H_\text{rh})$ plane and the constraints could be tighter if all the theoretical uncertainties listed below were resolved.
In this sense, our scenario is {both highly predictive and} falsifiable.
Two major uncertainties are governed by purely SM physics:
(i) the temperature dependence of the Weinberg angle at the EW crossover [Eq.~\eqref{eq:f}] and (ii) the MHD evolution of the helical gauge fields after inflation with hierarchical Reynolds numbers and in the presence of a chiral asymmetry.
The first uncertainty (i) is related to the efficiency of baryogenesis. Its reduction would shrink the area of the green shaded region in Fig.~\ref{fig:etaB}. 
The resolution of the second one (ii) would give more robust upper (red) and lower (blue) bounds in Fig.~\ref{fig:etaB}.
The other two uncertainties are linked to the inflation sector:
(iii) the backreaction from the induced current and (iv) the reheating phase.
Since the backreaction from the production of SM fermions makes the system highly non-linear, we just give two estimates corresponding to the left and right panels in Fig.~\ref{fig:etaB}.
A rigorous treatment of this system requires a full numerical simulation treating not only the gauge field but also the SM fermions dynamically, which would give a solid prediction of the hyper gauge field as a function of $(\xi, H_\text{inf})$.
Finally, throughout this paper, we have assumed instantaneous reheating.
Although a recent study~\cite{Cuissa:2018oiw} indicates that reheating is completed immediately for a moderate value of the Chern-Simons coupling (see footnote \ref{ft:reheating}), a regime with a smaller $\xi$ parameter may be affected significantly if the inflaton instead enters a prolonged oscillation phase.

%%%%%%%%%%%%%%%%%%%%%%%%%%%%%%%%%%%%%%%%%%%%%%%%%%%%%%%%%%%%%

\vspace{1cm}
\noindent \textbf{Acknowledgements} \newline
We thank Ruth Durrer, Tomohiro Fujita, Kohei Kamada, Kai Schmitz, Jennifer Schober and Lorenzo Ubaldi for insightful discussions. We moreover thank Kai Schmitz and Kohei Kamada for pertinent remarks on the manuscript. This work was funded by the Deutsche Forschungsgemeinschaft under Germany's Excellence Strategy - EXC 2121 ``Quantum Universe'' - 390833306.
EM would like to acknowledge the Mainz Institute for Theoretical Physics (MITP) of the DFG Cluster of Excellence PRISMA+ (Project ID 39083149) for its hospitality and its partial support during the completion of this work.

%%%%%%%%%%%%%%%%%%%%%%%%%%%%%%%%%%%%%%%%%%%%%%%%%%%%%%%%%%%%%

%\clearpage
\appendix

\section{Chemical equilibrium}
\label{sec:chemical}
Here we briefly summarize various relations for the chemical potentials resulting from the SM interactions before and after the EW phase transition. We also provide solutions for these chemical equations by identifying the slowest process for each regime.

\vspace{1em}
\paragraph*{Before the EW phase transition.}
We have $16$ different chemical potentials: $(\mu_{L_i},\mu_{e_i}, \mu_{Q_i},\mu_{u_i},\mu_{d_i},\mu_H)$ for $i = 1,2,3$.
The relevant SM interactions yield the following relations among them:
\begin{itemize}
	\item EW sphaleron: $0 = \sum_i \left( 3 \mu_{Q_i} + \mu_{L_i} \right)$.
	\item strong sphaleron: $0 = \sum_i \left( 2 \mu_{Q_i} - \mu_{u_i} - \mu_{d_i} \right)$.
	\item lepton Yukawa: $0 = \mu_{L_i} - \mu_H - \mu_{e_i}$.
	\item up-type Yukawa: $0 = \mu_{Q_i} + \mu_H - \mu_{u_j}$.
	\item down-type Yukawa: $0 = \mu_{Q_i} - \mu_H - \mu_{d_j}$.
\end{itemize}
Under these SM interactions, we have four conserved quantities: U$(1)_Y$ and $B/3 - L_i$. Assuming that there are no initial asymmetries for them, we require:
\begin{itemize}
	\item U$(1)_Y$: $0 =  3 \left( \mu_{Q} + 2 \mu_{u} - \mu_{d} \right) - \sum_i \left( \mu_{L_i} + \mu_{e_i} \right) +  2\mu_H$.
	\item $B/3 - L_i$: $0 = 2 \mu_{Q} + \mu_{u} + \mu_{d} - 2 \mu_{L_i} - \mu_{e_i}$.
\end{itemize}
Here we have taken common chemical potentials for quarks in different generations because of the Yukawa couplings which link the generations, \textit{i.e.}, $\mu_Q = \mu_{Q_i}$, $\mu_u = \mu_{u_i}$, and $\mu_{d} = \mu_{d_i}$.

The slowest process in this regime is the electron Yukawa interaction. Hence one may study the chemical equilibration of this system conveniently by assuming that all the SM interactions except for the electron Yukawa are efficient, which leads to $15$ independent equations imposed on the chemical potentials.
Therefore we can express all the chemical potentials in terms of the right-handed electron one:
\begin{align}
		&\mu_{e_2} = \mu_{e_3} = \frac{7 \mu_{e_1}}{481}\,, \quad \mu_{L_1} =  -\frac{415 \mu_{e_1}}{962}\,, \quad \mu_{L_2} = \mu_{L_3} = \frac{59 \mu_{e_1}}{962}\,, \nonumber \\[.5em]
	&\mu_{Q} = \frac{33 \mu_{e_1}}{962}\,, \quad \mu_{u} = \frac{3 \mu_{e_1}}{37}\,, \quad \mu_{d} = - \frac{6 \mu_{e_1}}{481}\,, \quad \mu_H = \frac{45 \mu_{e_1}}{962}\,.
	\label{eq:summary}
\end{align}

\vspace{1em}
\paragraph*{After the EW phase transition.}
After the EW phase transition, the SU$(2)_W$ multiplets split and hence each component can have a different chemical potential, namely $\mu_{L_i} \to (\mu_{\nu_i}, \mu_{E_i})$, $\mu_{Q} \to (\mu_{U}, \mu_{D})$, and $\mu_H \to (\mu_{W_+}, 0)$.
Due to the CKM mixing, the chemical potentials for quarks in different generations are again equal which we immediately impose here to avoid unnecessary complications.
The relevant SM interactions in this regime yield the following relations:
\begin{itemize}
	\item EW sphaleron: $0 = \frac{1}{2} \sum_i \left( 3 \mu_{U} + 3 \mu_{D} + \mu_{\nu_i} + \mu_{E_i} \right)$.
	\item strong sphaleron: $
	0 = \sum_i \left( \mu_{U} + \mu_D - \mu_{u} - \mu_{d} \right)$.
	\item lepton Yukawa: $0 = \mu_{E_i} - \mu_{e_i}$,\, $0 = \mu_{\nu_i} - \mu_{W_+} - \mu_{e_i}$.
	\item up-type Yukawa: $0 = \mu_{U} - \mu_{u}$,\, $0 = \mu_{D} + \mu_{W_+} - \mu_{u}$.
	\item down-type Yukawa: $0 = \mu_{D} - \mu_{d}$,\, $0 = \mu_{U} - \mu_{W_+} - \mu_{d}$.
	\item charged current: $0  = \mu_{\nu_i} - \mu_{W_+} - \mu_{E_i}$, \, \\ \hspace*{2.4cm} $0  = \mu_{D} + \mu_{W_+} - \mu_{U}$.
\end{itemize}
From the conserved quantities, one gets:
\begin{itemize}
	\item U$(1)_\text{em}$: \\ $0 =  3 \left( 2 \mu_{U} - \mu_{D} + 2 \mu_{u} - \mu_{d} \right) - \sum_i \left( \mu_{E_i} + \mu_{e_i} \right) +  6\mu_{W_+}$.
	\item $B/3 - L_i$: $0 = \mu_{U} + \mu_D + \mu_{u} + \mu_{d} - ( \mu_{E_i} + \mu_{\nu_i} +  \mu_{e_i} )$.
\end{itemize}
The EW sphaleron becomes the slowest process in this regime because its reaction rate immediately drops after the EW phase transition.
We have $14$ different chemical potentials with $13$ independent relations plus the one from the EW sphaleron.
Hence, in the regime where all the SM interactions except for the EW sphaleron are efficient, we can again express all the chemical potentials in terms of the chemical potential of the right-handed electron:
\begin{align}
	&\mu_{E_i} = \mu_{e_i} = \mu_{e_1}\,, \quad
	\mu_U = \mu_u = \frac{9 \mu_{e_1}}{11}\,, \quad \mu_{D} = \mu_d = \frac{8 \mu_{e_1}}{11}\,, \nonumber \\[.5em]
	&\mu_{\nu_i} = \frac{12 \mu_{e_1}}{11}\,, \quad \mu_{W_+} = \frac{\mu_{e_1}}{11}\,.
	\label{eq:summary2}
\end{align}

\section{Scaling in the turbulent and viscous regimes}
\label{sec:reynolds}

Here we sketch how to estimate the dynamics of MHD by balancing the terms in the following governing equations [see Eqs.~\eqref{eq:B_field} and \eqref{eq:velocity}]:
\begin{align}
	\frac{\partial \bm{B}_Y}{\partial \eta} &= \frac{\bm{\nabla}^2}{\sigma_Y} 
	 \bm{B}_Y
	+ \bm{\nabla} \times \left( \bm{v} \times \bm{B}_Y \right)
	+ \frac{2 \alpha_Y}{\pi} \frac{\mu_{Y,5}}{ \sigma_Y} \bm{\nabla} \times \bm{B}_Y\,, \label{eq:magnetic_app} \\
	\frac{\partial}{\partial \eta} \bm{v} &= \nu \bm{\nabla}^2 \bm{v} - \bm{v} \cdot \bm{\nabla} \bm{v}  + \frac{1}{\rho + P}\left( - \frac{1}{2} \bm{\nabla} \bm{B}_Y^2 +( \bm{B}_Y \cdot \bm{\nabla}) \bm{B}_Y \right) \,. \label{eq:velocity_app}
\end{align}
In the following, we omit the third term in Eq.~\eqref{eq:magnetic_app} proportional to $\mu_{Y,5}$, which leads to the chiral plasma instability as will be discussed in the next Appendix~\ref{sec:cpi}.
The electric conductivity and the kinetic diffusion are estimated as
\begin{align}
	\sigma_Y = \frac{c_\sigma}{\alpha_Y \ln \alpha_Y^{-1}} T \,,
	\quad
	\nu = \frac{c_\nu}{\alpha_Y^2 \ln \alpha_Y^{-1}} \frac{1}{T}\,.
\end{align}

The dynamics of the hyper magnetic field is determined by the relative size between the diffusion term, $\bm{\nabla}^2 \bm{B}_Y / \sigma_Y$, and the induction term, $\bm{\nabla} \times (\bm{v} \times \bm{B}_Y)$,
which is characterized by the magnetic Reynolds number:
\begin{align}
	R_m = \sigma_Y L v\,.
\end{align}
Here $L$ is the typical length scale of the hyper magnetic field and $v$ is the typical amplitude of the velocity field.
If $R_m < 1$, the hyper magnetic field is dissipated away exponentially, and this is not the case we are interested in.
%which case is out of our interest.

Similarly, the dynamics of the velocity field is also characterized by the relative size between the diffusion term, $\nu \bm{\nabla}^2 \bm{v}$, and the induction term, $\bm{v} \cdot \bm{\nabla} \bm{v}$,
\textit{i.e.}, by the kinetic Reynolds number:
\begin{align}
	R_e = \frac{L v}{\nu}\,.
\end{align}
The ratio between $R_m$ and $R_e$ is the so-called magnetic Prandl number:
\begin{align}
	P_m = \frac{R_m}{R_e} \sim \alpha_Y^{-3} \gg 1\,.
\end{align}
Thus the magnetic Reynolds number is much larger than the kinetic one in the SM plasma at high temperatures.

Given that we thus have $R_m > R_e$, we can consider three regimes: (i) $R_m > R_e > 1$, (ii) $R_m > 1 > R_e$, (iii) $1 > R_m > R_e$.
The last case is not of our interest because then both the hyper magnetic field and velocity field are diffused away immediately.
In the following, we consider the first and second case which are referred to as the turbulent and viscous regimes, respectively.

\vspace{1em}
\paragraph*{Turbulent regime: $R_m > R_e > 1$.}
$R_m > 1$ implies that magnetic diffusion is inefficient.
Hence, the helical charge is approximately conserved:
\begin{align}
	\text{const.} = h \sim L B_Y^2\,.
	\label{eq:h_cnsv_app}
\end{align}
For $R_e > 1$, the amplitude of the velocity field can be estimated by balancing the source term from the hyper magnetic field and the induction term, $\bm{v}\cdot \bm{\nabla} \bm{v}$, in Eq.~\eqref{eq:velocity_app}.
This consideration leads to the equipartition of energy densities
\begin{align}
	\rho \, v^2 \sim B_Y^2\,.
	\label{eq:equi_app}
\end{align}
Inserting Eq.~\eqref{eq:equi_app} into Eq.~\eqref{eq:magnetic_app}, omitting magnetic diffusion because of $R_m > 1$ and using Eq.~\eqref{eq:h_cnsv_app}, one finds
\begin{align}
	\partial_\eta B_Y \sim \frac{v  B_Y}{L} \propto B_Y^4\,.
\end{align}
From this, we can estimate the scalings of relevant quantities in the turbulent regime as follows:
\begin{align}
	B_{Y,\eta} \sim \left( \frac{\eta_t}{\eta} \right)^\frac{1}{3} B_{Y, \eta_t}\,, \quad
	L_\eta \sim \left( \frac{\eta}{\eta_t} \right)^\frac{2}{3} L_{\eta_t}\,,
	\quad
	v_\eta \sim  \left( \frac{\eta_t}{\eta} \right)^\frac{1}{3} v_{\eta_t}\,,
	\label{eq:scaling_turb}
\end{align}
for $\eta > \eta_t$ with $\eta_t$ being the onset of the turbulence.
Once this scaling is achieved, both the magnetic and kinetic Reynolds numbers increase with time:
$R_m, R_e \propto v L \propto \eta^{1/3}$.

\vspace{1em}
\paragraph*{Viscous regime: $R_m > 1 > R_e$.}
While we still have helicity conservation \eqref{eq:h_cnsv_app} because of $R_m > 1$, the diffusion term dominates over the induction term in Eq.~\eqref{eq:velocity_app} for $R_e < 1$.
In this case, the amplitude of the velocity field can be estimated by balancing the source term from the hyper magnetic field with the diffusion term $\nu \bm{\nabla}^2 \bm{v}$:
\begin{align}
	v \sim \frac{L B_Y^2}{\rho \, \nu} \quad \leftrightarrow \quad
	\rho \, v^2 \sim R_e B_Y^2\,.
	\label{eq:velocity_viscous}
\end{align}
One can see that the equipartition no longer holds for $R_e < 1$.
Inserting Eq.~\eqref{eq:velocity_viscous} into Eq.~\eqref{eq:magnetic_app} and using Eq.~\eqref{eq:h_cnsv_app}, one finds
\begin{align}
	\partial_\eta B_Y \sim \frac{v B_Y}{L} \propto B_Y^3\,.
\end{align}
From this, we then get
\begin{align}
		B_{Y,\eta} \sim \left( \frac{\eta_t}{\eta} \right)^\frac{1}{2} B_{Y, \eta_t}\,, \quad
	L_\eta \sim \left( \frac{\eta}{\eta_t} \right) L_{\eta_t}\,,
	\quad
	v_\eta \sim v_{\eta_t}
	\label{eq:scaling_vis}
\end{align}
for $\eta > \eta_t$.
Again, both the magnetic and kinetic Reynolds numbers increase with time, $R_m, R_e \propto \eta$, once the system falls into this scaling.

\section{Chiral plasma instability}
\label{sec:cpi}

In this section we derive and solve the equations that govern the evolution of the helicity and of the energy density of the hyper gauge fields, together with the chemical potential $\mu_{Y,5}$, which, as discussed above, controls the asymmetries stored in all the fermion species.
Lacking, for the initial conditions of interest here, a full MHD simulation including both the effect of the chiral plasma and of the plasma velocity field, we will check the validity of our estimate Eq.~(\ref{eq:TCPI}) for the time scale of the chiral plasma instability by solving numerically the MHD equations neglecting the fluid velocity. Once Eq.~(\ref{eq:TCPI}) is verified, we will assume that its validity extends to the viscous and turbulent regime, whose only effect would be that of suppressing the effect of diffusion by means of the inverse cascade described above. In the main text, we combine this estimate of $\hat T_\text{CPI}$ with the helicity conservation found in the inverse cascade regime.
It would be certainly important to check the validity of this assumption, and we hope our work will motivate further studies in this direction.

We start from the Maxwell equations and the generalized Ohm's law, which, in a FLRW Universe with zero net hyper-charge, read
\begin{align}
&\partial_\eta {\bm E}_Y - {\bm\nabla}\times{\bm B}_Y + {\bm J}_Y = 0 \\
&\partial_\eta {\bm B}_Y + {\bm\nabla}\times{\bm E}_Y = 0 \\
&{\bm\nabla} \cdot {\bm E}_Y = 0 \\
&{\bm\nabla} \cdot {\bm B}_Y = 0 \\
&{\bm J}_Y =  \sigma_Y \left({\bm E}_Y + {\bm v}\times{\bm B}_Y\right) + \frac{2\alpha_Y}{\pi} \mu_{Y,5} {\bm B}_Y \label{eq:Ohm} \,.
\end{align}
In the following we will ignore the velocity-dependent term in Eq.~(\ref{eq:Ohm}).
In the MHD limit ($\partial_\eta^2{\bm B}_Y\approx 0$, $\partial_\eta{\bm E}_Y\approx 0$), one obtains
\begin{align}
&\partial_\eta {\bm B}_Y - \frac{1}{\sigma_Y} \left({\bm\nabla}^2{\bm B}_Y + \frac{2\alpha}{\pi} \mu_{Y,5} {\bm\nabla}\times{\bm B}_Y \right) = 0 \label{eq:B_appendix}\\
&{\bm E}_Y = \frac{1}{\sigma_Y} \left({\bm\nabla}\times{\bm B}_Y - \frac{2\alpha_Y}{\pi} \mu_{Y,5}{\bm B}_Y \right)\,.\label{eq:E_appendix}
\end{align}
Let us now define the hyper helicity density and the hyper magnetic energy density as
\begin{align}
h & = \frac{1}{\text{vol}\, (\mathbb{R}^3)} \int \dd^3 x\, {\bm A_Y}\cdot{\bm B_Y}\,, \\
\rho_B & = \frac{1}{\text{vol}\, (\mathbb{R}^3)} \int \dd^3 x\, \frac{1}{2} {\bm B_Y}^2 \,,
\end{align}
where
\begin{align}
{\bm B}_Y & = {\bm\nabla}\times{\bm A}_Y = \int\frac{\dd^3k}{(2\pi)^3}\sum_{\lambda=\pm} {\bm\nabla}\times\left({\bm\epsilon}_{\bm k}^\lambda A_Y^\lambda (\eta,{\bm k}) e^{i \bm{k} \cdot \bm{x}}\right) \nonumber \\
& = \int\frac{\dd^3k}{(2\pi)^3}\sum_{\lambda=\pm} \left(\lambda \,k \, {\bm\epsilon}_{\bm k}^\lambda A_Y^\lambda (\eta,{\bm k}) e^{i \bm{k} \cdot \bm{x}}\right) \nonumber \\
& \equiv \int\frac{\dd^3k}{(2\pi)^3}\sum_{\lambda=\pm} \left( {\bm\epsilon}_{\bm k}^\lambda B_Y^\lambda (\eta,{\bm k}) e^{i \bm{k} \cdot \bm{x}} \right) \,.
\end{align}
The transverse polarization tensor satisfies the relations
${\bm k}\cdot{\bm\epsilon}_{\bm k}^\pm = 0$,
$({\bm\epsilon}_{\bm k}^\lambda)^*\cdot{\bm\epsilon}_{\bm k}^{\lambda'} = \delta_{\lambda\lambda'}$, ${\bm k}\times{\bm\epsilon}_{\bm k}^\pm = \mp i k{\bm\epsilon}_{\bm k}^\pm$ and $({\bm\epsilon}_{\bm k}^\pm)^* = {\bm\epsilon}_{-{\bm k}}^\pm$.
Moreover, to ensure the reality of ${\bm A}_Y$ and ${\bm B}_Y$, we impose $A_Y^\lambda (\eta,{\bm k})^* = A_Y^{\lambda} (\eta,-{\bm k})$.
If we now define
\begin{align}
h & = \int \dd k\, h_k \,, \\
\rho_B & = \int \dd k\, \rho_{B,k} \,,
\end{align}
we obtain
\begin{align}
h_k & = \frac{k^3}{2\pi^2} \left(|A_Y^+ (\eta,{\bm k})|^2 - |A_Y^- (\eta,{\bm k})|^2\right) \,, \\
\rho_{B,k} & = \frac{k^4}{4\pi^2} \left(|A_Y^+ (\eta,{\bm k})|^2 + |A_Y^- (\eta,{\bm k})|^2\right) \,,
\end{align}
and
\begin{align}
\partial_\eta h_k & = -\frac{2k^2}{\sigma_Y}h_k + \frac{8\alpha_Y}{\pi}\frac{\mu_{Y,5}}{\sigma_Y} \rho_{B, k} \,, \label{eq:hk evolution}\\
\partial_\eta \rho_{B, k} & = -\frac{2k^2}{\sigma_Y}\rho_{B,k} + \frac{2\alpha_Y}{\pi}\frac{\mu_{Y,5}}{\sigma_Y} k^2 h_k \,,\label{eq:rhok evolution}
\end{align}
In both of the above equations, the first term on the right-hand side is due to diffusion, and leads to the exponential decay of the hyper magnetic field for a finite hyper electric conductivity. The second term is instead responsible for the chiral plasma instability, as we will discuss in the following.

The evolution of the chemical potential is governed by Eq.~(\ref{eq:anomaly_w_Ye}),
\begin{align}
\frac{ \partial \mu_{e_1}}{\partial \eta} & = - \frac{3}{T^2}\frac{\partial q_{\text{CS},Y}}{\partial \eta} - \Gamma_{Y_e} \frac{711}{481} \mu_{e_1} \nonumber \\
& = - \frac{3}{T^2}\frac{\alpha_Y}{\pi}\frac{\partial h}{\partial \eta} - \Gamma_{Y_e} \frac{711}{481} \mu_{e_1} \, ,
\label{eq:mu evolution}
\end{align}
where we recall that the comoving temperature $T$, defined as $T = a\,\hat T$, is constant as long as the expansion is adiabatic.
{When all the Yukawa couplings but the electron one are in equilibrium,
\begin{equation}\label{eq:mu5mue}
\mu_{Y,5} = \frac{711}{481} \, \mu_{e_1} \,.
\end{equation}
This is not true at high temperature, as the other Yukawa couplings progressively go out of equilibrium. For the purpose of this appendix, we will nevertheless assume it to simplify the discussion. Our main motivation will be to validate the estimate of $\hat T_\text{CPI}$ in Eq.~(\ref{eq:TCPI}) and ensure that the chiral plasma instability does not occur before the electron Yukawa goes into equilibrium, $\hat T_\text{CPI} \gtrsim 10^5\GeV$. In this regime Eq.~(\ref{eq:mu5mue}) is valid.
}
The initial condition is obtained, at the end of inflation, from the anomaly equation [see Eq.~(\ref{eq:mu_5_ye})]:
\begin{align}
\label{eq:mu h initial}
\mu_{e_1} = -\frac{3}{T^2} \frac{\alpha_Y}{\pi} h \, .
\end{align}
From Eq.~(\ref{eq:mu evolution}) it is clear that this condition is satisfied until the Yukawa interactions become important, which happens around $\hat T \sim 10^5\GeV$.

Let us now discuss the evolution of $h_k$ and $\rho_{B, k}$. To understand how the chiral plasma instability happens, it is interesting to define the helicity ratio $r_k$,
\begin{align}
r_k = \frac{k \, h_k}{2\rho_{B,k}}\,,
\end{align}
in such a way that $r_k=\pm 1$ for a maximally helical mode. Eqs.~(\ref{eq:hk evolution}) and (\ref{eq:rhok evolution}) yield
\begin{align}
\label{eq:rk}
\frac{\partial r_k}{\partial \eta} = \frac{4 \alpha_Y}{\pi}\frac{\mu_{Y,5}}{\sigma_Y} \, k \, (1-r_k^2) \,, \\
{\frac{\partial h_k}{\partial \eta} = \frac{2 k}{\sigma_Y} \left( \frac{k_\text{CPI}}{r_k} - k \right) h_k \,, }\label{eq:hk_rk_app} 
\end{align}
{with $k_\text{CPI} = 2 \alpha_Y \mu_{Y,5}/\pi$.}
The above equation shows that, for $\mu_{Y,5}<0$, $r_k = -1$ is a stable fixed point, while $r_k=+1$ is unstable (and vice versa for $\mu_{Y,5}>0$). Let us assume, without loss of generality, that the total initial helicity $h$ is positive, so that the initial chemical potential $\mu_{Y,5}$ is negative according to Eqs.~(\ref{eq:mu5mue}) and \eqref{eq:mu h initial}.
Consider a mode $k$ whose initial helicity is positive $h_k>0$ and quasi-maximal ($1-r_k\ll 1$). Neglecting the time variation of $\mu_{Y,5}$, the helicity ratio will initially decrease, cross 0 and then grow (in absolute value) asymptotically to 
$-1$.

Because of the factor $k$ in the right hand side of Eq.~(\ref{eq:rk}), the evolution will be faster for short wave-length modes, thus resulting in an inverse cascade behavior, in which power is transferred from short to long wave-lengths.

The process is suppressed for $|\mu_{Y,5}/\sigma_Y|\ll 1$, or equivalently for small helicity $h$. For this reason, in the presence of $\mu_{Y,5} \neq 0$, the system will tend to erase $h$ and $\rho_B$ faster than in the case of pure diffusion.\footnote{This conclusion is the opposite with respect to what was highlighted in~\cite{Boyarsky:2011uy}, in which the inverse cascade regime preserves the magnetic fields from decaying due to diffusion. The reason of this discrepancy is the relative sign of $h$ and $\mu_{Y,5}$. When these quantities have the same sign, the maximally helical configuration is stable, and the helicity is sustained by the chiral plasma.}

To check the correctness of our analytical understanding of the onset of the chiral plasma instability, we solve numerically Eqs.~(\ref{eq:hk evolution}), (\ref{eq:rhok evolution}) and (\ref{eq:mu evolution}), and compare the result with Eq.~(\ref{eq:TCPI}).
Once diffusion becomes efficient, the hyper magnetic field is washed out and the instability becomes insignificant. In reality, the onset of the turbulent or viscous regime (tied to the velocity field of the plasma which we are omitting in this appendix) will prevent diffusion from becoming relevant. For the purpose of this appendix, we will simply choose initial conditions such that $\hat T_\text{CPI}$ is larger than the temperature at which diffusion becomes efficient. The latter can be estimated from the diffusion term in Eq.~(\ref{eq:hk evolution}) as
\begin{align}
\hat T_\text{diff} = \frac{M_\ast}{T}\frac{1}{\eta_\text{diff}} \sim \frac{M_\ast}{T}\frac{2 \bar k^2}{\sigma_Y} \,,
\end{align}
where $\bar k$ is the comoving mode that carries most of the helicity, and we recall that $M_\ast \equiv [90 / (\pi^2 g_\ast)]^{1/2} \Mpl$.

In the absence of fermion backreaction, the enhanced super-horizon hyper gauge fields scale as $A_{Y,-}(k) \propto 1/\sqrt{k}$ [see Eq.~\eqref{eq:gauge_desitter}]. Consequently
the helicity spectrum generated at the end of inflation scales as $h_k\propto k^2$ and has a UV cut-off at $k_\text{UV} \sim (a H)_\text{rh} = 
 H_\text{inf} T/ \hat T_\text{rh}$.
We assume that fermion production does not affect the spectral behavior, but only the normalization, which is obtained from (cf. Eq.~(\ref{eq:qcs0}))
\begin{align}
h_\text{rh} \simeq - \frac{2}{3}\left(\frac{T}{\hat T_\text{rh}}\right)^3 \frac{\langle \hat{\bm E}_Y \cdot \hat{\bm B}_Y \rangle_\text{rh}}{H_\text{inf}} \,.
\end{align}

The chiral plasma instability happens when the second term on the right-hand side of Eqs.~(\ref{eq:hk evolution}) and (\ref{eq:rhok evolution}) dominates over the first. Our initial conditions are such that $\mu_{Y,5}\propto h$. Moreover, for a (almost) maximally helical field, $\rho_{B,k} \propto h$, so that the second term is $\propto h^2$ while the first is only $\propto h$. For this reason we expect the chiral effects to be more important when the total helicity is larger.

\begin{figure}[t]
\centering
\includegraphics[width=.45\textwidth]{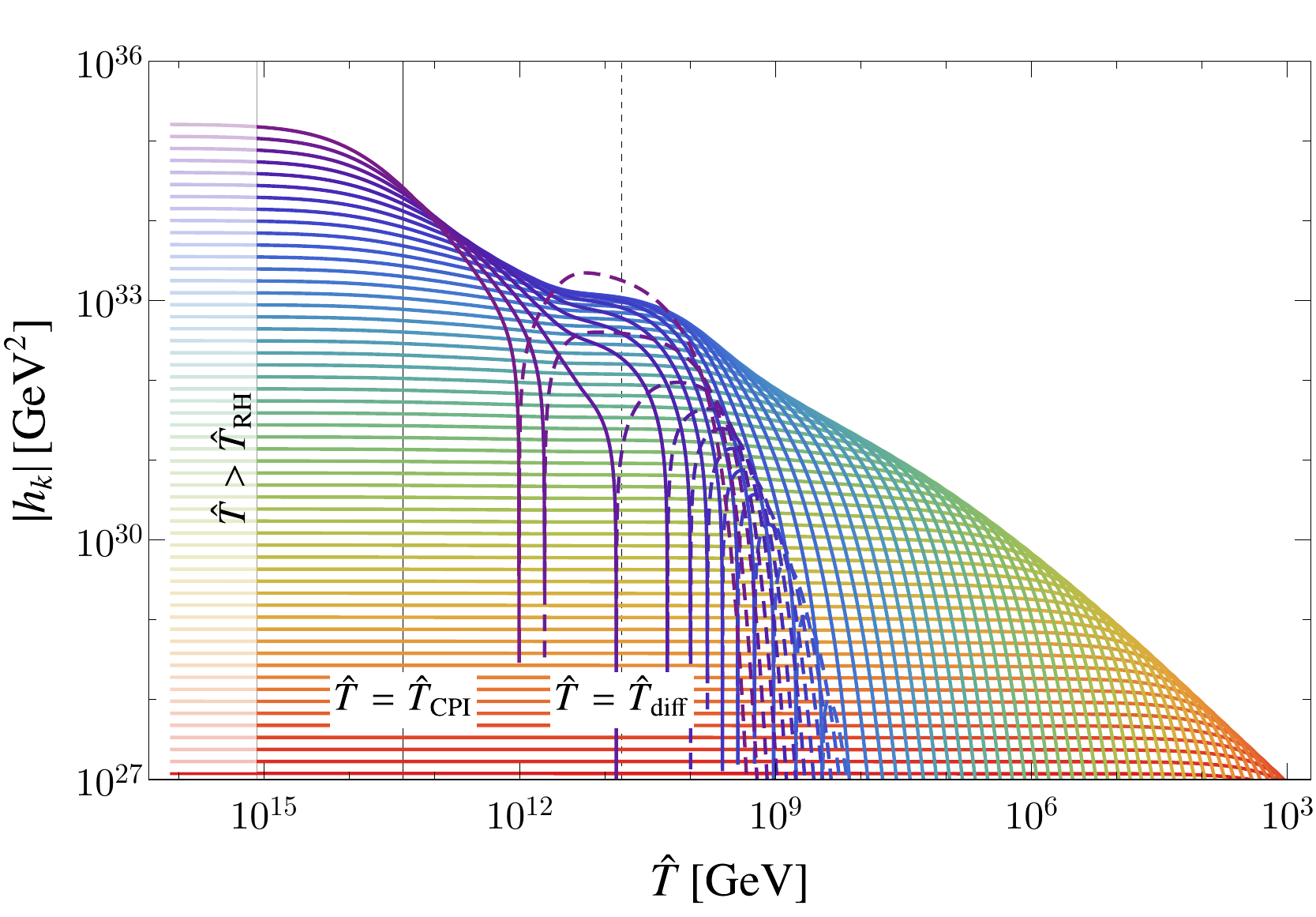}
\caption{Evolution of the helicity $h_k$ of logarithmically spaced Fourier modes, ranging from $10^{-4}~k_\text{UV}$ (red) to $k_\text{UV} \sim (a H)_\text{rh}$ (purple). Dashed lines indicate negative values for $h_k$. For pedagogical reasons we extend the plot to temperatures even above the instant reheating temperature, to clearly show the conservation of helicity at early times.} 
\label{fig:cpi hk}
\end{figure}
\begin{figure}[t]
\centering
\includegraphics[width=.45\textwidth]{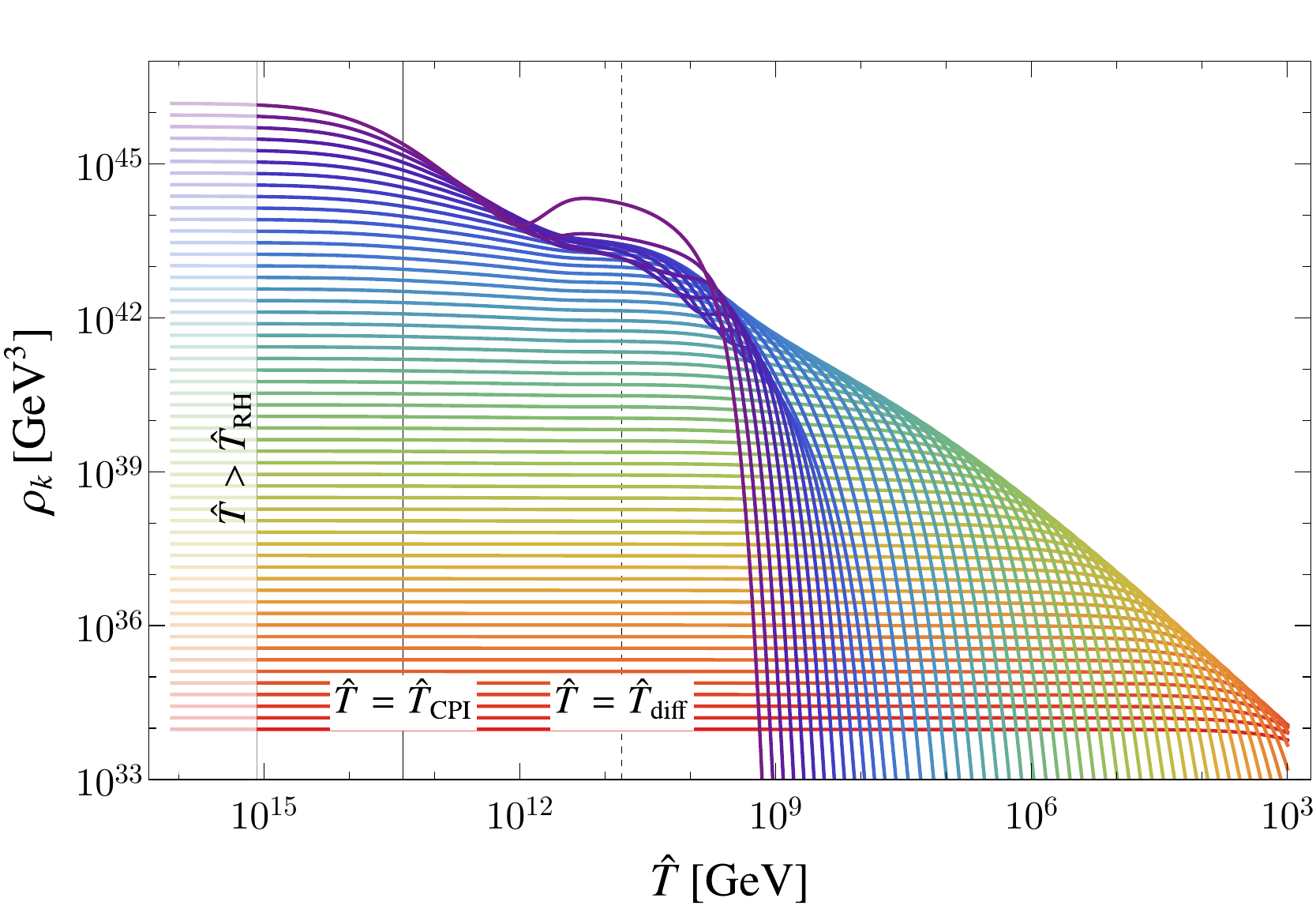}
\caption{Evolution of the energy density $\rho_k$, color-coding as in Fig.~\ref{fig:cpi hk}.}
\label{fig:cpi rhok}
\end{figure}
\begin{figure}[t]
\centering
\includegraphics[width=.45\textwidth]{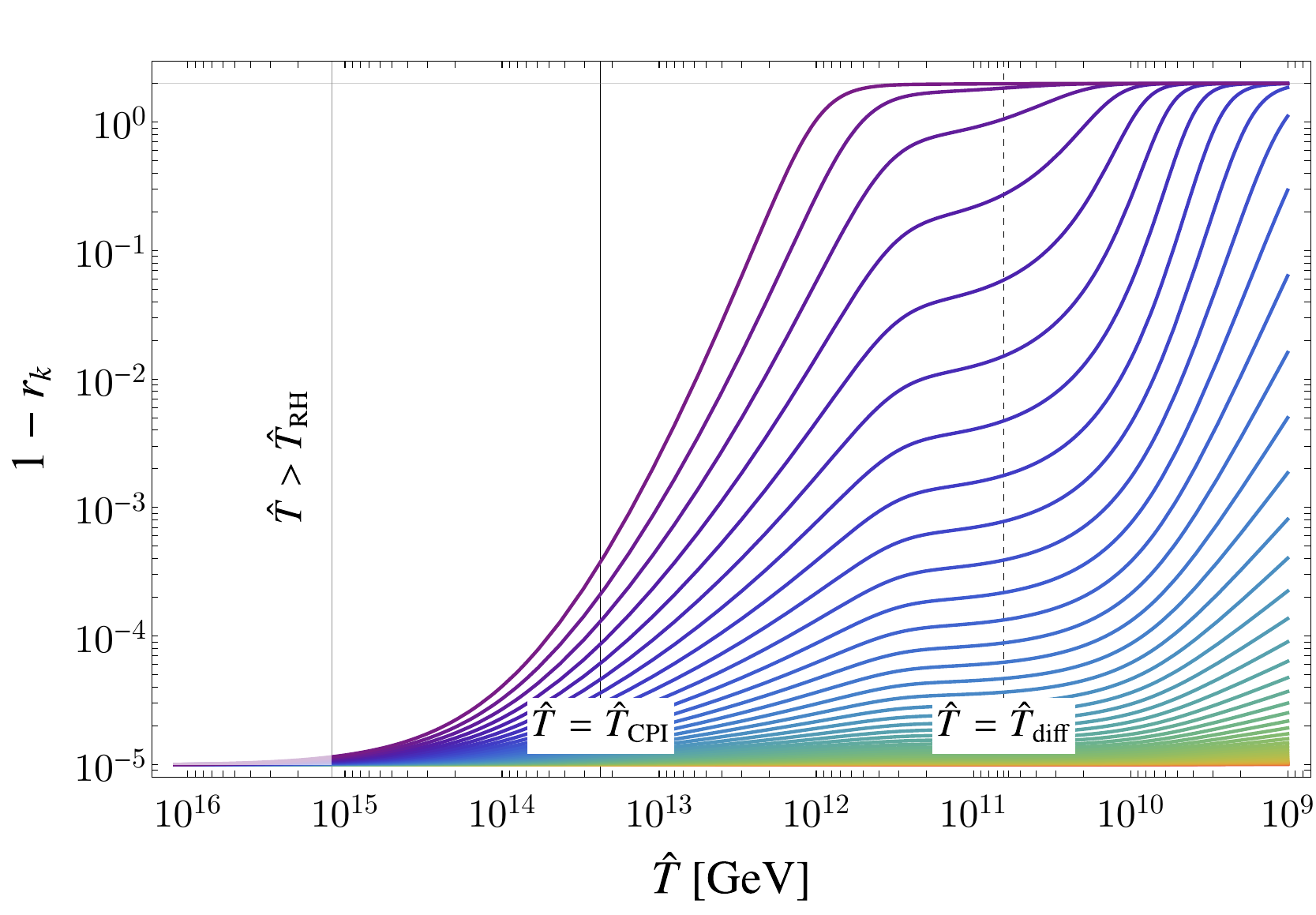}
\caption{Evolution of $r_k$, denoting the degree of helicity. Modes which are initially nearly maximally helical with a positive helicity, $r_k \sim 1$, are driven to the maximally negative helicity configuration, $r_k = -1$.}
\label{fig:cpi rk}
\end{figure}

\begin{figure}[t]
\centering
\includegraphics[width=.45\textwidth]{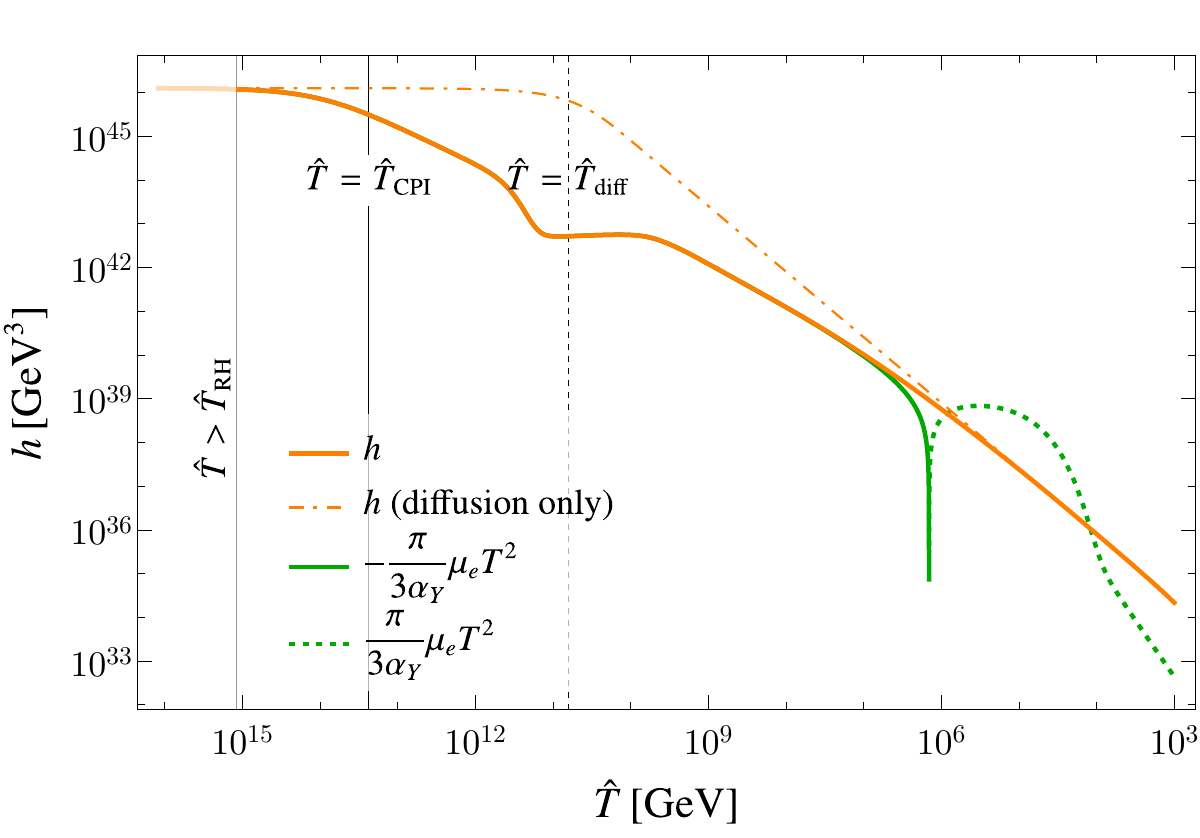}
\caption{Evolution of the total helicity $h$ and the right-handed electron chemical potential $\mu_{e_1}$. This demonstrates the conservation law~\eqref{eq:mu h initial} as well as its violation when the electron Yukawa reaches equilibrium and erases the electron chemical potential. For reference, the dashed dotted orange line indicates the evolution of the total helicity in the absence of chiral asymmetries.}
\label{fig:cpi mu}
\end{figure}

We show in Figs.~\ref{fig:cpi hk} to \ref{fig:cpi mu} the result of a numerical evaluation of Eqs.~(\ref{eq:hk evolution}), (\ref{eq:rhok evolution}) and (\ref{eq:mu evolution}), from $\hat T_\text{rh} \approx 10^{15}\GeV$ down to  $\hat T \approx 10^{3}\GeV$.
In order to make the chiral plasma instability clearly visible in the plots, we choose an artificially large value for the hyper gauge fields at the end of inflation, $\hat E \hat B = 3\times 10^{12} H_\text{inf}^4$. We evaluate $\alpha_Y$ at the scale $(\hat{E}\hat{B})^{1/4}$ at the end of inflation, $\alpha_Y\approx 0.015$.
We take $H_\text{inf} = 10^{-6} \Mpl$, $\sigma_Y \approx 70 T$ and, following~\cite{Bodeker:2019ajh}, $\Gamma_{Y_e} \approx 9.2\times 10^{-14} T $. 
We include $55$ modes, logarithmically spaced, defined as $k_n / k_\text{UV} = \exp[\log(2) \, (n-55)/4 ]$, with $k_\text{UV}$ defined above.
Finally, we take $T = 10^{14}\GeV$, but we remark that this choice does not affect the physics as it just rescales all the comoving quantities.

The evolution of the Fourier components of the helicity and energy density (see Figs.~\ref{fig:cpi hk}, \ref{fig:cpi rhok} and \ref{fig:cpi rk}) as well as the total helicity and the right-handed electron chemical potential (Fig.~\ref{fig:cpi mu}) clearly feature four distinct regions: (i) the conservation of all these (comoving) quantities for $\hat T \gg \hat T_\text{CPI} >  \hat T_\text{diff}$, (ii) the dynamics associated with the plasma instability around $\hat T \sim \hat T_\text{CPI}$, (iii) the dissipation of energy into the SM plasma at $\hat T < \hat T_\text{diff}$ and finally (iv) the re-generation of the a chemical potential with opposite sign driven by the (diffusion induced) decay of the total helicity at $\hat T \sim 10^5$~GeV. We note that taking into account the velocity field of the SM plasma, the latter effect will be highly suppressed since we expect diffusion to be much less efficient.

Let us turn in more detail to point (ii), which is the main focus of this appendix. For $h_k > 0$, $\mu_{Y,5} < 0$ both terms on the right-hand side of Eq.~\eqref{eq:hk evolution} are negative. The time scale of the resulting decrease in $h_k$ depends on the value of $k$ and occurs earlier for the UV modes of the spectrum (purple lines in Figs.~\ref{fig:cpi hk} to \ref{fig:cpi rk}). Let us first consider the most UV mode of the spectrum. Once its decreasing helicity crosses zero it continues to decrease (hence growing in absolute value), until $r_k = -1$ is saturated (maximally negative helical configuration) and the right-hand sides of both Eqs.~\eqref{eq:hk evolution} and \eqref{eq:rhok evolution} become zero. This indicates a stationary solution for $h_k$ and $\rho_{B,k}$. 
Since the most UV mode was initially the main carrier of the total helicity, this entire process leads to a significant decrease of the total helicity and, according to Eq.~\eqref{eq:mu h initial}, of $\mu_{Y,5}$. This implies that for all other modes, the onset of the chiral instability is `delayed', visible in a plateau-like feature in Figs.~\ref{fig:cpi hk} and \ref{fig:cpi rhok} at $\hat T \gtrsim \hat T_\text{CPI}$. Finally, at $\hat T \sim \hat T_\text{diff}$ diffusion becomes relevant (in particular for the most UV mode), dissipating the helicity of this mode and restarting a similar process for the subsequent mode. 

Some comments are in order.  First, in deriving Eq.~\eqref{eq:TCPI} we have used that the dominant mode driving the plasma instability is $k \sim k_\text{CPI}$. In this appendix, due to our particular choice of initial condition explained above, and in particular due to the initial large value of $\hat E \hat B$, $k_\text{CPI}$ is not covered by the range of modes in our numerical simulation, $k_\text{CPI} > k_\text{UV}$. The vertical line labeled $\hat T_\text{CPI}$ in Figs.~\ref{fig:cpi hk} to \ref{fig:cpi mu} thus corresponds to Eq.~\eqref{eq:TCPI} after replacing $k_\text{CPI} \mapsto k_\text{UV}$ as the dominant mode driving the chiral plasma instability. We see in Fig.~\ref{fig:cpi hk} that this estimate nicely coincides with the observed drop of $h_{k_\text{UV}}$, denoted by the top most purple line. We also note that the erasure of the helicity is not an instantaneous process (see also Fig.~\ref{fig:cpi mu}), indicating that the region close to the red line in Fig.~\ref{fig:etaB} deserves a more detailed analysis.

Second, since for any mode with $|r_k k|\ll |k_{\rm CPI}|$, the diffusion term in Eq.~\eqref{eq:hk_rk_app} is negligible and the CPI term then leads to a phase with growing $|h_k|$, one might worry about UV modes with $r_k \sim 0$ which are not covered in our numerical simulation. However, from Eq.~\eqref{eq:rk} it follows that once $|h_k|$ starts to grow, $|r_k|$ increases too. For any mode with $k > |k_{\rm CPI}|$, the diffusion term in Eq.~\eqref{eq:hk_rk_app} then eventually becomes important again and leads to the decay of that mode. This generalizes the conclusion discussed below Eq.~\eqref{k_CPI}. Note also that as long as such UV modes are bounded by the thermal spectrum, $\rho_{B,k} \lesssim k^2 T$, they would in any case not contribute significantly to the total energy or helicity density, and hence could not erase the chiral charge of the system.
The condition that the total energy or helicity density changes significantly turns out to be equivalent to $k \lesssim k_\text{CPI}$.

Third, we note that in the absence of the diffusion term, the stationary solution of Eqs.~\eqref{eq:hk evolution} and \eqref{eq:rhok evolution} would correspond to $\mu_{Y,5} = 0$, \textit{i.e.}, to a complete erasure of any helical and chiral charge. On the contrary, the finite diffusion ensures that $|\mu_{Y,5}| \simeq \pi k/(2 \alpha_Y) \neq 0$, with $k$ denoting the time-dependent UV cutoff of the (diffusion dampened) spectrum. This indicates that the balance between the chiral plasma instability and the overproduction of  the baryon asymmetry in the yellow region of Fig.~\ref{fig:etaB} might open a second window of viable baryogenesis at slightly larger values of $H_\text{rh}$. A more detailed study of this phenomenon is beyond the scope of the present paper.

%%%%%%%%%%%%%%%%%%%%%%%%%%%%%%%%%%%%%%%%%%%%%

\bibliography{refs2}{}

\end{document}